\documentclass[12pt,preprint]{aastex}

\newcommand{\kms}{km s$^{-1}$ }
\newcommand{\schi}{{\sc Hi}\ }
\newcommand{\schii}{{\sc Hii}\ }
\newcommand{\lv}{($\ell$,$v$) }
\newcommand{\lb}{($\ell$,$b$) }
\newcommand{\cmc}{cm$^{-3}$ }

\newcommand{\ms}{M$_{\odot}$ }
\newcommand{\vlsr}{$v_{\rm LSR}$ }
\newcommand{\vdev}{$v_{\rm dev}$ }

\shorttitle{\schi 21-cm Forbidden-Velocity Wings}
\shortauthors{J.-h. Kang, B.-C. Koo} 

\begin{document}

\title{ Faint \schi 21-cm Emission Line Wings \\
	at Forbidden-Velocities}

\author{Ji-hyun Kang and Bon-Chul Koo}
\affil{Astronomy Program, Department of Physics and Astronomy, Seoul National
University, Seoul 151-747, KOREA}

\email{kjh@astro.snu.ac.kr; koo@astrohi.snu.ac.kr}

\begin{abstract}

We present the results of a search for faint \schi 21-cm emission line wings
at velocities forbidden by Galactic rotation in the Galactic plane
using the Leiden/Dwingeloo \schi Survey data
and the \schi Southern Galactic Plane Survey data.
These ``forbidden-velocity wings (FVWs)'' appear
as protruding excessive emission in comparison with their surroundings
in limited ($\la 2 ~\degr$) spatial regions over velocity extent
more than $\sim 20$ \kms in large-scale \lv diagrams.
Their high-velocities imply
that there should be some dynamical phenomena associated.
We have identified 87 FVWs.
We present their catalog, and discuss their distribution and statistical properties.
We found that 85\% of FVWs are not coincident
with known supernova remnants (SNRs), galaxies, or high-velocity clouds.
Their natures are currently unknown.
We suspect that many of them are fast-moving \schi shells and filaments 
associated with the oldest SNRs that are essentially invisible
except via their \schi line emission.
We discuss other possible origins.

\end{abstract}


\notetoeditor{}

\section{Introduction}

\begin{figure}
\plotone{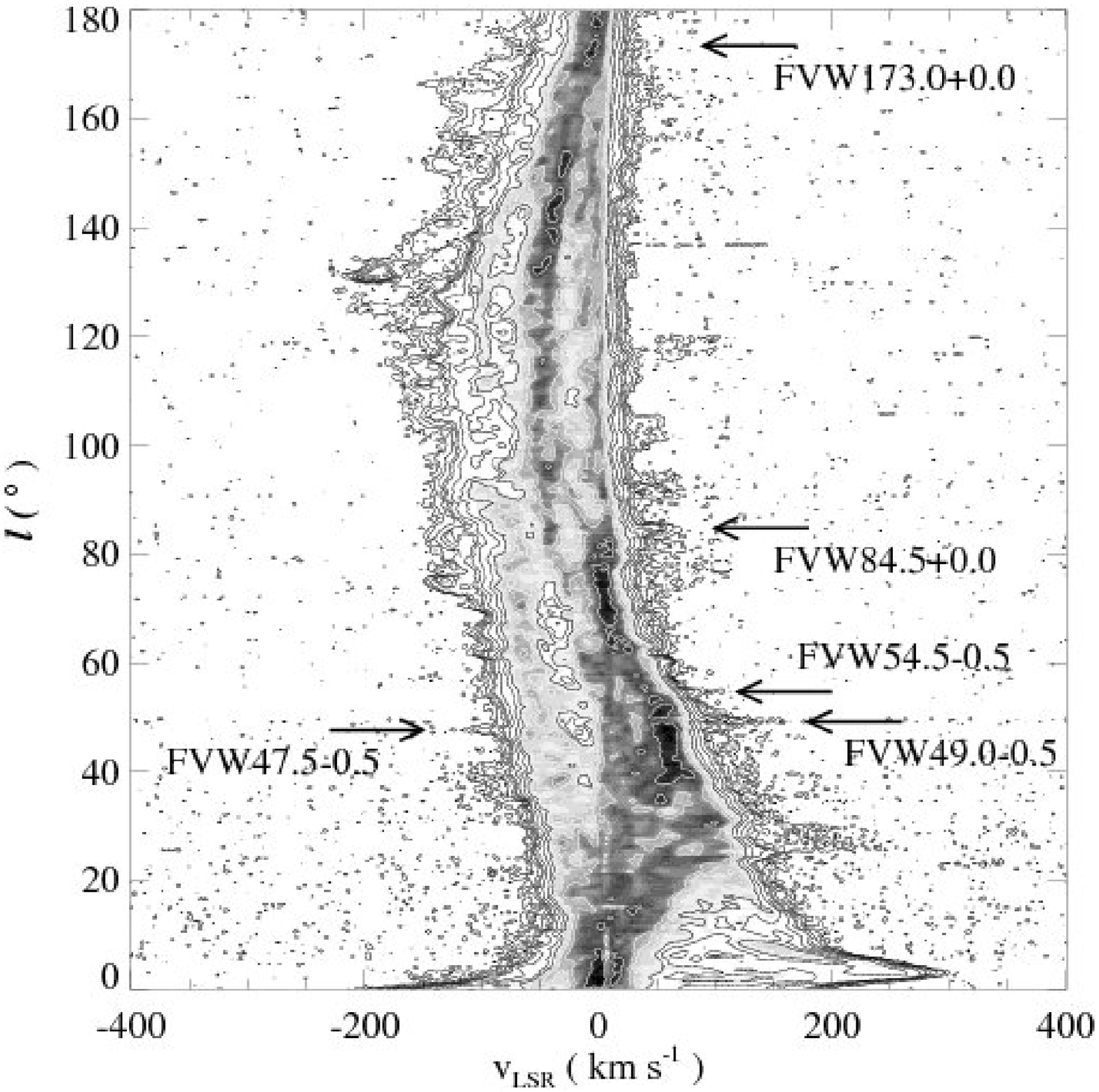}
\figcaption{Large-scale \lv diagram of the 1st and 2nd quadrants at $b=-0.\degr5$.
The LDS data are used.
Contours are 0.1, 0.2, 0.4, 0.7, 1, 2, 5, 10, 30, 50, 100~K in brightness temperature.
\label{introfig}}
\end{figure}

Large-scale \lv diagrams of \schi 21-cm line emission
in the Galactic plane usually show faint high-velocity bumps
protruding from their surroundings.
These faint ``wing"-like features are extended
to the velocities well beyond the maximum or minimum velocities permitted
by the Galactic rotation.
Some of those ``forbidden-velocity wings (FVWs)" are shown in Fig.~\ref{introfig},
which is a \lv diagram of the Galactic \schi emission
in the 1st and the 2nd quadrants at $b=-0\degr.5$.
FVWs are the ones that are small and smoothly extend
from the main Galactic \schi,
which distinguishes them from high-velocity clouds (HVCs)
that have discrete peaks isolated from the Galactic \schi.
FVWs are probably the sites where kinetic energies are expelled into the interstellar medium (ISM) by some violent events.

Our interest about FVWs was first motivated by the statistics of SNRs,
namely that the number of known SNRs is much less than that of the expected
\citep[][hereafter KK04]{kk04}. 
The number of presently-known Galactic 
SNRs, most of which have been identified in radio continuum,  
is 265 according to the Green's catalog \citep{green06} 
\footnote{Available at "http://www.mrao.cam.ac.uk/surveys/snrs/"},
which provides a most 
complete and up-to-date list of identified Galactic SNRs. 
This is much less than the number 
(20 -- 30,000) expected from the SN rate and the life time of SNRs.
Therefore, most SNRs are ``missing'', 
and it must be mainly old SNRs that are missing considering their large population.
This is not surprising because old SNRs are faint in radio continuum, and 
it is difficult to identify faint radio sources because of 
the confusion due to the Galactic background emission 
and also because of observational limitations \citep[e.g.,][]{brogan06}. 
KK04 paid attention to the fact that
an old SNR consists of fast expanding \schi shell
and that the shell might last after the remnant
becomes too faint to be visible in radio continuum.
If its expansion velocity is greater than the minimum or maximum
velocities permitted by the Galactic rotation,
then the SNR shell or part of it (maybe the caps) could be detected as high-velocity gas,
e.g., FVWs, in Fig.~\ref{introfig}. They proposed that the FVWs could be
possible candidates for the missing old SNRs.
Indeed, \citet{kks06} have carried out high resolution \schi line observation
toward FVW 190.2+1.1, one of FVWs in this paper,
and detected a rapidly expanding ($\sim 80$~\kms) \schi shell.
The parameters of this shell seem only consistent with
those of the remnant of a SN explosion.
This shell is not seen in any other wave band,
suggesting that it represents the oldest type of SNR,
that is essentially invisible except via its \schi line emission.

Although SNR is a strong candidate of FVWs,
the high velocities of FVWs could be also produced by stars
expelling their mass into the ISM,
e.g., stellar winds from massive stars, and
outflows from young stellar objects or asymptotic giant branch (AGB) stars.
High- or intermediate-velocity clouds (HVCs, IVCs)
that happen to be close to the maximum/minimum velocity
of the \schi gas along the line of sight would appear as FVWs, too.
In order to understand the nature of FVWs and their relation
to old SNRs,
therefore, a systematic study of FVWs is required.

We have searched FVWs in the Galactic plane ($\mid b \mid \le 13\degr$)
using Leiden/Dwingeloo \schi Survey (LDS) data and
the \schi Southern Galactic Plane Survey (SGPS) data,
and studied their statistical characteristics.
Some of our preliminary results were published by \citet{kkh04}.
In Section~\ref{secdata}, we explain the \schi surveys
used for the search and the identification procedure.
The catalog of FVWs
is presented in Section 3,
where we show their statistical properties and the result of comparison with
other objects, too.
Possibility of FVWs to be SNRs and some other possible origins
will be discussed in Section 4.
We finally summarize our results in Section 5.

\section{Observational Data and Identification\label{secdata}}

\subsection{Data}
We used two Galactic \schi survey data; the Leiden/Dwingeloo \schi survey (LDS)
\citep{hb97} and the Parkes survey,
which is a part of the \schi Southern Galactic Plane Survey (SGPS)
\citep{mc01}.

The LDS
was carried out with the 25-m Dwingeloo telescope
(FWHM $= 36^\prime$)
on a half-degree grid
using frequency-switching mode.
The survey covered the sky north of declination $-30\degr$
completely on this grid.
Its effective velocity coverage spans
Local Standard of Rest (LSR) velocities from $-450$ to $+400$~\kms,
resolved into spectral channels of 1.03~\kms width.
Stray radiation correction is applied to LSD data.
The nominal brightness-temperature sensitivity of the LDS is 0.07~K.
The velocity resolution is not critical for the identification of FVWs 
because FVWs are extended over a quite wide velocity range.
We smoothed each spectrum to a resolution of 3.09 \kms, 
so that the rms noise of the final cubes used for the identification 
becomes $\sim 0.05$ K. 

The Parkes survey is a part of SGPS, a survey of the \schi spectral line
and 21-cm continuum emission in the fourth quadrant of the Galactic plane.
The Parkes survey covers $l=253\degr$--$358\degr$ and $|b|\la 10\degr$. 
The Parkes data were obtained with the inner seven beams
of the Parkes multibeam system,
a 13 beam 21-cm receiver package at prime focus
on the Parkes 64 m radio telescope (FWHM$ = 16^\prime$), 
by the process of ``on-the-fly'' mapping.
The final calibrated Parkes data consist of 10 cubes
with a cell size of $4^\prime$,
and a velocity resolution of 0.82 \kms
covering from $-250$ to $+200$~\kms.
The rms noise of the survey in brightness temperature
varies from 0.13 to 0.27~K depending on the cube.
The main survey area ($\mid b \mid \le 1\degr .5$)
was re-observed to put limits on the amount of stray radiation,
so that the rms noise is significantly better in this area ($0.09 \sim 0.14$~K).
For the coherence of this study, we convolved the Parkes data
using a Gaussian beam
to have the same angular resolution as the LDS
and then re-gridded the data with a grid interval of $0\degr .5$.
The velocity channels are not smoothed (but just regridded to a 
grid of 1.03~\kms\ interval) because the resulting data cubes have 
rms noise (0.02--0.04 K) already lower than that of the 
LDS data cubes. 

\subsection{Identification}

We have identified FVWs by drawing
large scale \lv diagrams of each
quadrant every $0\degr .5$ in Galactic latitude
for Galactic latitudes of $\mid b \mid \le 13\degr$. 
All identification was carried out with bare eyes. 
We first selected areas
protruding from the Galactic background \schi emission
at contours between 0.1~K and about 1.0~K.
Among them, we chose the ones with velocity extent
greater than $\sim$20~\kms compared to the neighborhood area.
Then, we looked into the channel images to make sure that
their high-velocity features are clearly visible.
We limited our selection only for those that are confined in small regions
($\lesssim 2\degr$).
The $\la 2$~\degr criterion is not based on an astrophysical consideration,
but is rather empirical. However, this should be a reasonable choice for
identifying missing SNRs because most of them might reside in the inner Galaxy
and smaller than $\sim 2$~\degr (KK04).
In Fig.~\ref{introfig}, for example, 
the features near $\ell = 47\degr, 49\degr,54\degr$,
$84\degr$ and 173\degr are selected.
But, the ones confined
in somewhat broad longitude regions ($\sim 5\degr $)
near $\ell=27\degr$, 40\degr, and 100\degr
are not selected.
Those features do not seem to be real
but to be caused by data reduction process.
In LDS data, the spectra in a $5\degr \times 5\degr$ box
are reduced at same time and 
these excess emissions are
confined in $5\degr \times 5\degr$ square areas almost exactly.

FVWs could not be identified
in the areas where groups of HVCs or intermediate-velocity clouds (IVCs), 
or supershells make the velocity boundary complicated
or where the velocity boundary is unclear or clumpy.
For example, Anti-center shell, and HVC complexe H and G
highly contaminate the negative velocity boundaries
near $120\degr \la \ell \la 200\degr$,
and some velocity boundaries at high latitudes in the inner Galaxy
are clumpy due to many HVCs.
Table~\ref{comarea} summarizes those areas.
We note that there are gaps at
$ 242\degr \lesssim \ell \lesssim 253\degr$,
at $0\degr \la \ell \la 5\degr$, and at $350\degr \la \ell \la 360\degr$
between the two surveys.
(The exact longitude interval depends on latitude.)

\clearpage
\begin{deluxetable}{cccl}
\tabletypesize{\scriptsize}
\tablewidth{0pt}
\tablecaption{Highly Complicated Areas in ($\ell, v$) Diagrams\label{comarea}}
\tablehead{
\colhead{($\ell_{\rm min}, \ell_{\rm max}$)} &
\colhead{($b_{\rm min}, b_{\rm max}$)} & \colhead{\vlsr ($+$ or $-$)}      &
\colhead{Note} }
\startdata
(0\degr, 50\degr) & ($-13$\degr,  $-4$\degr) & $+$ & \\
(10\degr, 30\degr) & ($+5$\degr, $+10$\degr) & $+$ & \\
(10\degr, 30\degr) & ($+7$\degr, $+13$\degr) & $-$ & \\
(90\degr, 150\degr) & ($-11$\degr, $-7$\degr) & $-$ & HVC Complexe G\\
(110\degr, 160\degr) & ($-2$\degr, $+7$\degr) & $-$ & HVC Complexe H\\
(160\degr, 180\degr) & ($-2$\degr, $+4$\degr) & $-$ & Anti-Center Shell\\
(160\degr, 200\degr) & ($+4$\degr, $+13$\degr) & $-$ & Anti-Center Shell\\
(180\degr, 205\degr) & ($-4.5$\degr, $+0$\degr) & $-$ & Anti-Center Shell\\
(260\degr, 330\degr) & ($+5$\degr, $+10$\degr) & $+$& \\
(330\degr, 360\degr) & ($+6$\degr, $+10$\degr) & $+,-$& \\
(330\degr, 360\degr) & ($-10$\degr, $-4$\degr) & $+,-$& \\
\enddata
\end{deluxetable}
\clearpage

\section{Results}

\subsection{The Catalog}
\notetoeditor{We wish to make this catalog machine-readable table,
which is included as '\input{table}'
at the end of this subsection.
And, we'd like to show the entire table in paper, too.
}

Totally 87 FVWs have been identified;
73 in the northern sky
and 14 in the southern sky.
The parameters of the FVWs are summarized in Table~\ref{catafvw}.
The entries of the catalog are the following.: 
(1) Name of FVWs based on the Galactic coordinates.
The central coordinates are determined in the images integrated
over the given velocity extent.
(2) Velocity ranges where excess \schi emission is seen.
(3) Integrated intensity of \schi 21-cm line emission over the given velocity
extent at the center of FVWs.
(4) Rank indicating definiteness of protruding feature.
Higher rank means it is more apparent in \lv or ($b,v$) diagram
(see next paragraph).
(5) Associated celestial objects
(see Section 3.3).

We assign rank 3 to FVWs that are well-defined in both in \lv and ($b, v$)
diagrams. The prototypical ones are FVW 49.0$-$0.5, FVW 94.5+8.0, FVW 203.0+6.5,
etc. They are usually located in areas with flat and clear velocity
boundaries. Rank 1 FVWs are those with relatively unclear high-velocity
features in both \lv and ($b, v$) diagrams. The prototypical ones are
FVW 10.0+7.5, FVW 51.5+3.5, FVW 201.5$-$7.5, etc. They are located in
areas of complicated velocity boundaries, but still show stronger
emission than their surroundings over the velocity intervals in Table 2.
We assign rank 2 to the rest, i.e., the ones that are relatively
well-defined in \lv diagram but not in ($b, v$) diagram, or vice versa.
The prototypical ones are FVW 6.5$-$2.5, FVW 197.0+7.0, FVW 311.9$-$1.1.
Rank 3 includes 33 FVWs and both of rank 2 and rank 1 have 27 FVWs.
We note that, for some FVWs,
e.g., FVW $99.5-7.5$, FVW $104.5+2.0$, or FVW 196.5+1.5,
the classification is ambiguous
so that the ranks may be considered as a general guide of protrusion level.
The integrated images and position-velocity diagrams
of 87 FVWs are presented in Fig.~\ref{fvws}.
Since identifying FVWs by eyes is a subjective process, we note that
the FVW catalog may be incomplete and needs a careful consideration for
statistical interpretation (see next section).

\clearpage
\begin{deluxetable}{lcrcl}
\tablewidth{0pt}
\tabletypesize{\scriptsize}     
\tablecolumns{6}
\tablecaption{Catalog of Forbidden-Velocity Wings\label{catafvw}}
\tablehead{
\colhead{Name}  & \colhead{$v_{\rm min}$,$v_{\rm max}$} &
\colhead{$\int \Delta \bar{T_{b}} \,dv$}        & \colhead{}    &
\colhead{Associated}    \\
\colhead{(FVW$\ell \pm b$)}     &\colhead{(km s$^{-1}$)}       &\colhead{(K km s$^{-1}$)}     &
\colhead{Rank}  & \colhead{Object}}
\startdata
FVW6.5$-$2.5    &$-$135,$-$108&  3.3$\pm$  5.3  &2&\\
FVW10.0+7.5     & $-$81, $-$39& 28.6$\pm$  4.6  &1&\\
FVW15.5$-$10.5  &$-$48, $-$17   & 14.9$\pm$  1.6 &2&\\
FVW18.0$-$6.5   & $-$53, $-$23& 16.3$\pm$  2.4  &2&\\
FVW20.0$-$5.5   & $-$45, $-$24&  3.7$\pm$  3.1  &2&\\
\\
FVW27.5+6.5     &$-$101, $-$75& 10.0$\pm$  2.5  &2&\\
FVW34.0$-$1.5   & 117, 138      & 19.6$\pm$  8.4 &1&\\
FVW39.0+4.0     &  89, 131      & 35.2$\pm$  8.2 &2&\\
FVW40.0+0.5     &$-$113, $-$80& 21.9$\pm$  2.3  &3&\\
FVW44.5$-$2.0   &  79, 119      & 53.8$\pm$ 14.2  &1&\\
\\
FVW47.5$-$0.5   &$-$111, $-$88&  6.5$\pm$  0.8  &3&\\
FVW49.0$-$0.5   &  91, 138      & 34.9$\pm$  4.6  &3&SNR, W 51\\
FVW51.5+3.5     &  71,  86      & 24.3$\pm$  9.7  &1&\\   
FVW54.5$-$0.5   &  85, 105      &  2.8$\pm$  0.7  &2&SNR, HC 40\\
FVW69.0+2.5     &  35,  89      & 26.6$\pm$  5.6  &3&SNR, CTB 80\\
\\
FVW71.0$-$4.0   &$-$129,$-$103& 57.4$\pm$  5.1  &2&\\
FVW75.5+0.5     &  30,  60      & 29.9$\pm$  3.0 &1&\\
FVW78.0+2.0     &  43,  57      & 13.0$\pm$  3.5 &1&SNR, DR 4\\
FVW79.0+1.0     &  35,  57      & 22.4$\pm$  5.6 &1&\\
FVW81.5+1.5     &  33,  53      & 41.0$\pm$  5.2 &1&\\
\\
FVW83.5+4.0     &  30,  52      & 29.0$\pm$  5.7  &1&\\
FVW84.5+0.0     &  34,  55      & 45.9$\pm$  5.7  &1&\\
FVW86.0$-$9.5   &  20,  57      & 54.0$\pm$  7.7  &2&\\
FVW87.5$-$11.0  &  21,  42      & 12.0$\pm$  4.4  &1&\\
FVW88.5+5.0     &  30,  88      & 24.6$\pm$  5.2  &3&SNR, HB 21\\
\\
FVW94.5+8.0     &  31, 114      & 18.5$\pm$  2.2  &3&Galaxy, Cepheus 1\\
FVW95.5+11.5    &  25, 180      & 45.3$\pm$  2.4  &3&Galaxy, NGC 6946\\
FVW95.5+7.0     &  26,  65      &  7.3$\pm$  2.4  &3&\\
FVW96.0+4.0     &  18,  38      & 14.3$\pm$  2.5  &2&\\
FVW96.5$-$6.0   &  28,  48      &  7.5$\pm$  1.3  &1&\\
\\
FVW97.5$-$3.0   &  28,  52      &  5.5$\pm$  1.3  &1&\\
FVW99.5$-$7.5   &  11,  48      & 60.5$\pm$ 12.2  &2&\\
FVW101.0$-$9.5  &  20,  57      & 28.9$\pm$  4.8  &3&\\
FVW104.5+2.0    &  20,  52      &  7.4$\pm$  1.4  &3&\\
FVW107.0+8.0    &  17,  58      & 15.2$\pm$  3.0  &3&\\
\\
FVW108.0+7.0    &  27,  53      & 5.4$\pm$  2.0   &3&\\
FVW108.0+10.0   &  29,  51      &  3.2$\pm$  0.8  &3&\\
FVW109.5+2.5    &   8,  44      &264.8$\pm$ 40.3  &3&\\
FVW112.0$-$6.0  &  18,  36      & 11.7$\pm$  1.6  &1&\\
FVW112.0$-$2.0  &  39,  65      &  4.7$\pm$  0.8  &1&SNR (?), Cassiopeia A\\
\\
FVW116.5+12.0   &$-$196,$-$147& 14.0$\pm$  1.8  &3&\\
FVW126.0$-$1.0  &  24,  55      &  5.1$\pm$  0.9  &3&\\
FVW128.0$-$10.5 &   9,  58      & 50.3$\pm$  5.8  &2&\\
FVW133.5+2.5    &  26,  47      &  1.5$\pm$  0.6  &2&\\
FVW138.0+10.5   &  16, 129      &170.2$\pm$ 12.0  &3&Galaxy, UGC 2847\\
\\
FVW139.5+10.5   &  26, 155      & 32.7$\pm$ 52.5  &3&Galaxy, UGCA 86\\
FVW166.5+0.5    &  29,  59      &  4.7$\pm$  1.0  &2&\\
FVW172.5+6.0    &  28,  86      & 11.2$\pm$  1.1  &3&\\
FVW173.0+0.0    &  22,  40      &  5.5$\pm$  1.8  &2&\\
FVW173.0+3.0    &  11,  40      & 57.0$\pm$ 23.7  &3&\\
\\
FVW190.2+1.1    &  44,  81      &  8.4$\pm$  1.6  &3&\\
FVW191.0$-$2.0  &  54,  82      &  5.7$\pm$  1.4  &3&\\
FVW196.5+1.5    &$-$130, $-$58& 74.8$\pm$ 11.1    &1&HVC \\
FVW197.0+7.0    &  52,  78      &  4.4$\pm$  2.1  &2&\\
FVW201.0+9.5    &  64,  90      &  9.2$\pm$  0.8  &3&\\
\\
FVW201.5$-$7.5  & $-$56, $-$20& 11.8$\pm$  4.2    &1&\\
FVW201.5$-$5.0  & $-$60, $-$17& 17.6$\pm$  3.9  &1&\\
FVW203.0+6.5    &  85, 132      & 22.4$\pm$  1.3 &3&HVC \\
FVW203.5+4.5    &  80, 106      &  5.5$\pm$  2.3 &3&\\
FVW211.5+9.0    &  66,  98      & 10.7$\pm$  3.8 &1&\\
\\
FVW212.5+5.5    &  78, 115      &  9.8$\pm$  2.2 &3&\\
FVW212.5+3.5    &  67, 127      & 45.1$\pm$ 13.5 &2&\\
FVW215.0+7.0    &  84, 115      & 11.1$\pm$  1.7 &3&HVC\\
FVW215.5+7.5    & $-$43, $-$20&  5.8$\pm$  0.8  &3&\\
FVW225.5+6.0    & 103, 128      &  7.2$\pm$  2.1 &1&\\
\\
FVW233.5+5.0    & 114, 148      &  8.9$\pm$  1.6 &1&\\
FVW235.5$-$1.0  & 134, 155      &  4.0$\pm$  1.2 &1&\\
FVW238.5+0.5    & 141, 167      &  5.2$\pm$  0.9 &3&\\
FVW240.0+4.0    & 127, 153      &  7.6$\pm$  2.2 &1&\\
FVW242.5+2.0    & 141, 161      &  6.7$\pm$  1.7 &1&\\
\\
FVW243.0$-$2.0  & 135, 167      & 19.0$\pm$  3.5 &3&\\
FVW251.0+12.0   & $-$36, $-$13& 27.9$\pm$  4.2  &3&\\
FVW252.5+10.5   & $-$50, $-$30&  6.6$\pm$  0.7  &2&\\
FVW260.9$-$1.9  & 159, 184      & 16.8$\pm$  5.9  &2&\\
FVW261.6+2.3    & 158, 178      & 13.9$\pm$  5.9  &2&\\
\\
FVW283.4+3.0    & $-$57, $-$42& 13.0$\pm$  5.0  &2&\\
FVW291.0+2.9    & $-$72, $-$55& 17.9$\pm$  9.8  &2&\\
FVW293.0+7.9    & $-$74, $-$53& 20.5$\pm$  7.8  &1&\\
FVW298.3$-$3.7  & 135, 167      & 19.2$\pm$  8.4 &1&\\
FVW304.5$-$5.7  & $-$91, $-$66& 40.7$\pm$ 13.4  &2&\\
\\
FVW311.9$-$1.1  &$-$109, $-$92&  8.9$\pm$  4.1  &2&\\
FVW312.3+3.6    & $-$99, $-$80& 42.3$\pm$ 36.1  &2&\\
FVW313.4+0.6    &$-$117,$-$100& 12.2$\pm$  4.8  &2& \\
FVW315.2+2.7    &$-$107, $-$87& 20.9$\pm$  8.9  &1&\\
FVW319.8+0.3    &$-$123, $-$98& 22.9$\pm$  8.1  &3&\\
\\
FVW324.5+4.2    &$-$132,$-$112&  7.0$\pm$  4.8  &2&\\
FVW342.5+1.7    &  96, 110      &  3.9$\pm$  1.9        &3&\\
\enddata
\end{deluxetable} 
\clearpage

\subsection{Statistical Properties}

\begin{figure}
\epsscale{1.0}
\plotone{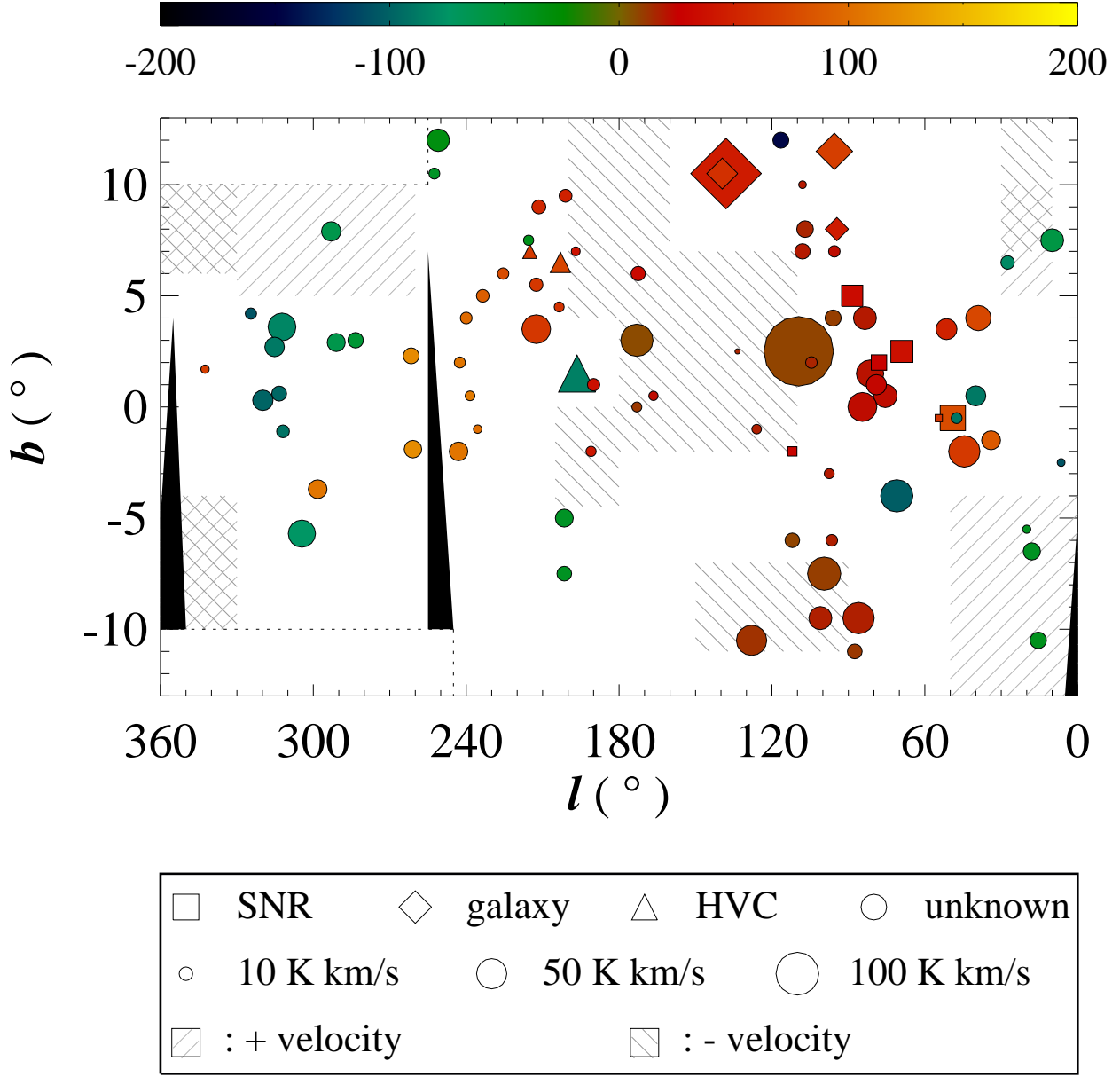}
\figcaption{Distribution of FVWs in \lb plane. The color of each symbol
indicates the mean velocity of the excess emission
and corresponds to the velocity in the color bar on the top in \kms.
Each symbol indicates related object,
and its occupied area indicates the integrated antenna temperature
as shown in the bottom.
The hatched areas represent 
the highly complicated areas in Table~\ref{comarea}.
The areas filled with solid black show gaps between data,
and the dotted line shows the boundary of the search in latitude.
\label{lbvifig}}
\end{figure}

Fig.~\ref{lbvifig} presents the distribution of the FVWs in \lb plane
with information about their LSR velocities and integrated intensities.
FVWs are marked by different symbols
depending on their associations with other known objects (see Section 3.3).
The center of each symbol indicates its Galactic coordinates
and the area presents its integrated intensity.
The color indicates the mean velocity with the scales shown by
the color bar on the top.
The hatched areas represent the highly complicated areas in Table~\ref{comarea}.
The areas filled with solid black show gaps between data,
and the dotted line shows the boundary of the search in latitude.

\begin{figure}
\epsscale{1.0}
\plottwo{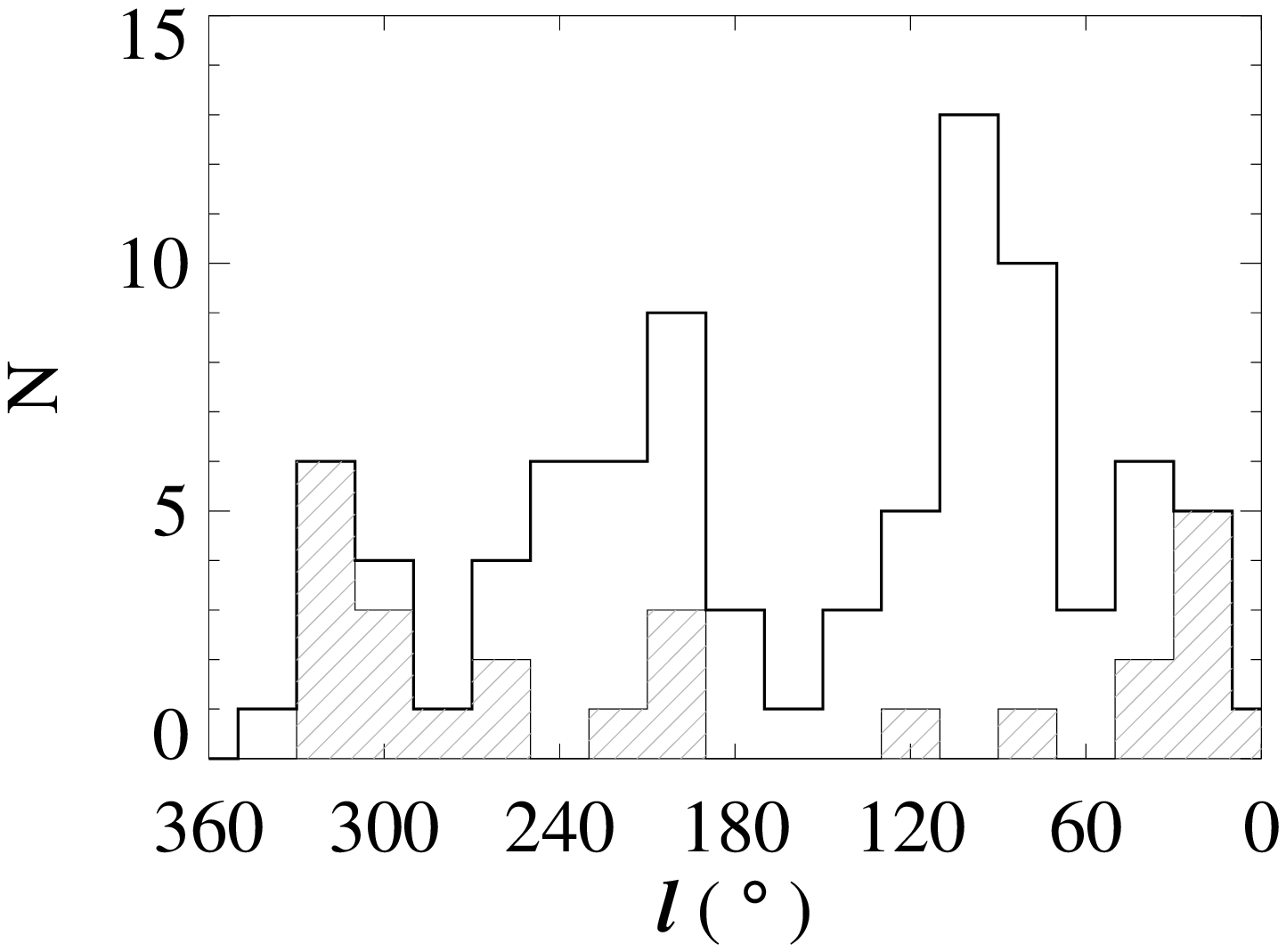}{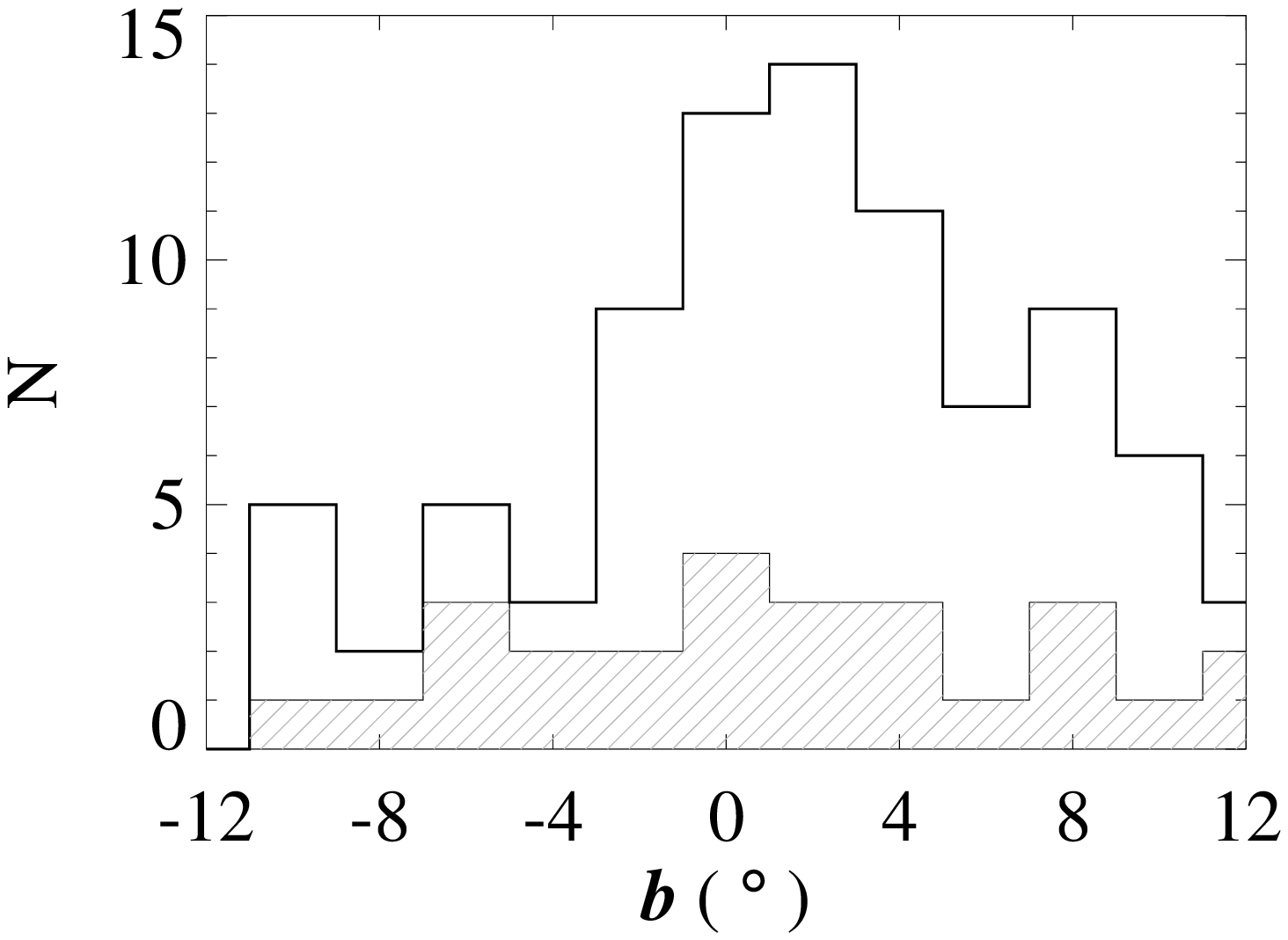}
\figcaption{(left) Distribution of FVWs in Galactic longitude.
The hatched area shows the FVWs at negative velocities,
while the white area indicates those at positive velocities.
(right) Same as the left but for Galactic latitude.\label{distlbn}}
\end{figure}

FVWs are scattered on the \lb plane.
They are neither clustered in the inner Galaxy
nor strongly concentrated toward the Galactic plane.
This is more clearly seen in Figs.~\ref{distlbn} (left) and (right)
which show the distribution of FVWs in Galactic longitude and latitude,
respectively.
Along the Galactic longitude, there are relatively large number of FVWs
between $\ell \approx 80\degr$ and $120\degr$
and perhaps at $\ell \approx 200\degr$, while 
the number is relatively small 
between $\ell \approx 140\degr$ and $180\degr$ and 
between $\ell \approx 330\degr$ and $10\degr$.
The distribution of FVWs in Galactic latitude is broad and 
slightly skewed to high positive latitudes. 
A Gaussian fitting yields $b=2.\degr 3$ as the center 
and $13.\degr 2$ as the full width at half maximum (FWHM) 
of the distribution. 
The FWHM in the inner Galaxy is smaller than that in the outer Galaxy, i.e., 
$\sim 8$ and 17$\degr$, respectively.

A noticeable feature in Fig.~\ref{lbvifig} is the change of the velocity of FVWs
with Galactic longitude (see also Fig.~\ref{distlbn} (left)).
Between $\ell \sim 50\degr$ and $250\degr$,
most (90~\%) FVWs have positive velocities,
while in the rest, FVWs with negative velocities are dominant (70~\%).
In total, the fraction of FVWs at positive velocities is
significantly greater, 70\% vs. 30\%.
The distribution of FVWs with positive and negative velocities are
also markedly different in their latitude distribution (Fig.~\ref{distlbn} (right)):
Positive-velocity FVWs are concentrated toward the Galactic plane
although the FWHM is large,
while negative-velocity FVWs are roughly uniformly distributed.

The above statistical properties might reflect 
the detectability of FVWs in the first place, e.g., 
how smooth and sharp the boundary of the general Galactic 
\schi emission is in ($\ell,v$) diagram. 
In the 1st and 2nd quadrants, 
the boundaries at negative-velocities
(hereafter negative-boundaries),  
which correspond to the edge of the Galaxy,
are wiggly and diffuse, so that the identification of 
negative FVWs are difficult there.
Also, anti-center shell (ACS) and HVC complexes make negative-boundaries in 
the 2nd quadrant complicated (Table~\ref{comarea}).
The boundaries at 
positive velocities
(hereafter positive-boundaries) 
are generally smooth and sharp, although they become 
complicated toward the inner Galaxy, particularly between
$\ell\approx 0\degr$ and $\sim 30\degr$ 
where there is not much ISM
and the gas motion is not circular.
The above explains why most of the FVWs in the 1st and 2nd quadrants 
have positive velocities and also why there are relatively few FVWs 
in the innermost Galaxy.
In the 3rd and 4th quadrants, it is negative-boundaries which are 
smooth and sharp. But they are 
not as sharp as the positive-boundaries in the 1st and 2nd quadrants.
In addition, ACS makes the negative-boundaries at 
$\ell\approx 180\degr$--$200\degr$ complicated.
This may explain why the negative-velocity FVWs in these quadrants 
are not as dominant as the positive-velocity FVWs in the 
1st and 2nd quadrants.
 
On the other hand, there is no reason for positive-velocity FVWs 
at $\ell=130\degr$--$180\degr$ 
to be more difficult to be identified 
than those at $\ell=80\degr$--$120\degr$ because both areas  
have very smooth and sharp positive-boundaries.
Therefore, there should be intrinsically less 
positive-velocity FVWs in this area.
Also there is no obvious changes in boundary shapes to cause 
the asymmetry and the broad extent in latitude distribution. 

\begin{figure}
\plotone{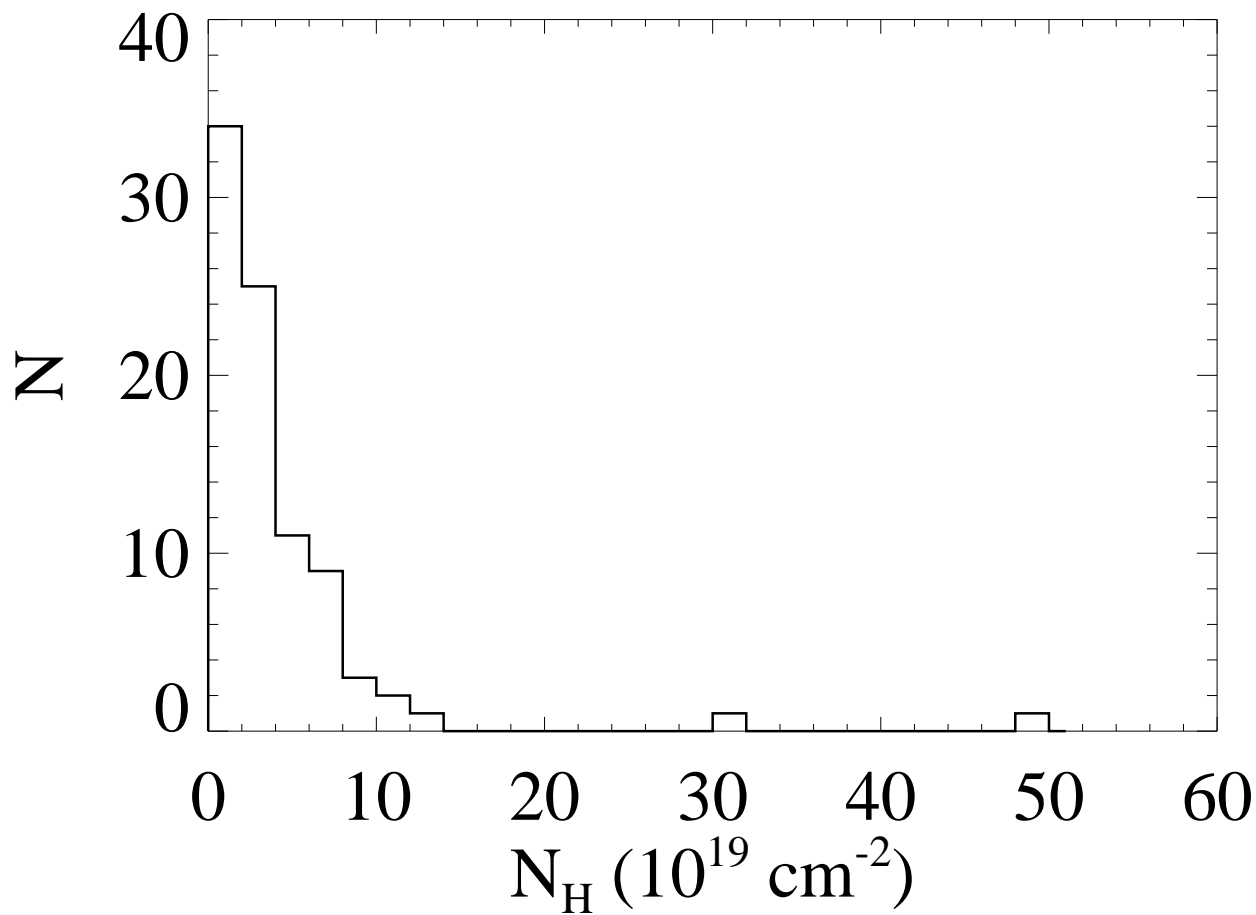}
\figcaption{Column density distribution of FVWs.\label{colden}}
\end{figure}

Fig.~\ref{colden} shows the distribution of column density of FVWs.
About 90~\% of FVWs have values
N$_{\rm H} \la 8 \times 10^{19}$~cm$^{-2}$.
The median N$_{\rm H}$ is $2.8 \times 10^{19}$~cm$^{-2}$.
Note that this is the column density over the velocity interval in Table~\ref{catafvw}.

\subsection{Comparison with Other Known Objects}
\clearpage
\begin{deluxetable}{lllccc}
\tabletypesize{\scriptsize}
\tablewidth{0pt}
\tablecaption{FVWs Coincident with SNRs\label{snrs}}
\tablehead{
\colhead{}    &  \multicolumn{4}{c}{SNR Parameters}\\
\cline{2-6} \\
\colhead{FVW Name}           & \colhead{Name}      &
\colhead{Other Name(s)}           & \colhead{Size ($\arcmin$)}      &
\colhead{Distance (kpc)}	&\colhead{Reference}}
\startdata
FVW49.0$-$0.5&G49.2$-$0.7& W51& 30&4.1 &1 \\
FVW54.5$-$0.5&G54.4$-$0.3& HC40& 40&3.3 &2 \\
FVW69.0+2.5&G69.0+2.7& CTB 80& 80&2.0&3\\
FVW78.0+2.0&G78.2+2.1& DR4, gamma Cygni SNR      & 60&1.4&2\\
FVW88.5+5.0&G89.0+4.7& HB21&120 $\times$ 90&0.8&4\\
FVW112.0$-$2.0&G111.7$-$2.1& Cassiopeia A, 3C461     &5  &3.4&5\\
\enddata
\tablerefs{
(1) \citet{sato73}; (2) \citet{milne79}; (3) \citet{koo90};
(4) \citet{uyaniker03}; (5) \citet{reedetal95}.}
\end{deluxetable}
\clearpage
We compared the positions of FVWs with those of SNRs, nearby galaxies,
HVCs, and found that 6, 4, and 3 FVWs are
coincident with those objects, respectively.

For comparison with SNRs, we used the
Green's SNR catalog \citep{green06}.
We have found six FVWs coincident with SNRs
and the results are summarized in Table~\ref{snrs}.
Systematic studies to search for high-velocity \schi gas
have been made toward 200 SNRs \citep{kh91, kkm04}.
They identified 25 SNRs with fast-moving \schi gas localized on SNR areas.
Five SNRs out of 25 are identified as FVWs in this paper, i.e., 
FVWs in Table~\ref{snrs} excluding FVW112.0$-$2.0
which is coincident with Cas A.
The other 20 SNRs are not identified as FVWs
due to several reasons.
Some SNRs' expansion velocities are not large
enough to be identified as a FVW
in a large-scale \lv diagram.
Some have small size so that their excess emission at high velocities
are smoothed out in the data with low angular resolution.
For three SNRs in Table~\ref{snrs},
high-resolution studies of the shocked \schi gas
have been made previously, e. g.,
W51 by \citet{km97},
CTB80 by \citet{koo90},
and DR4 by \citet{bs86}.
We have detected FVW toward Cas A, but,
considering that
Cas A is only $\sim 300$ years old,
it seems to be a chance coincidence.
Spectrum of Cas A region is very noisy
due to strong radio continuum.
\clearpage
\begin{deluxetable}{lrlrccc}
\tabletypesize{\scriptsize}
\tablewidth{0pt}
\tablecaption{FVWs Coincident with Nearby Galaxies\label{gals}}
\tablehead{
\colhead{}    &  \multicolumn{6}{c}{Galaxy Parameters}\\
\cline{2-7} \\
\colhead{} 	& \colhead{($\ell$, $b$)} &
\colhead{}           & 
\colhead{Size}      &\colhead{Velocity\tablenotemark{a}}&
\colhead{Distance}	&\colhead{}\\
\colhead{FVW Name}           & 
\colhead{(\degr)} &
\colhead{Other Name(s)}           & 
\colhead{($\arcmin$)}      &\colhead{(\kms)}&
\colhead{(Mpc)}	& \colhead{Reference}}
\startdata

FVW94.5+8.0 & (94.37, +8.01)&Cepheus 1& 11.7&58&6.0&1\\
FVW95.5+11.5& (95.72, +11.68)&NBG 2146, NGC 6946  &11.2&46&5.5&2\\
FVW138.0+10.5& (138.18, +10.58)&NBG 329, UGC 2847  &16.1&32&3.9&2\\
FVW139.5+10.5& (139.77, +10.64)&NBG 348, UGCA 86& 1.0&72&4.4&2\\

\enddata
\tablenotetext{a}{Heliocentric systematic velocity}
\tablerefs{
(1) \citet{burton99}; (2) \citet{tully}.}
\end{deluxetable}
\clearpage
For comparison with galaxies,
we used the `Nearby Galaxies Catalog' of \citet{tully},
which includes 2367 galaxies with systemic velocities less than 3000~\kms.
Three FVWs are identified as galaxies according to this catalog.
Another FVW, FVW94.5+8.0, looks similar to those 3 FVWs,
and turned out to be a low surface brightness galaxy named Cep 1
\citep{burton99}. Their parameters are summarized in Table~\ref{gals}.
FVWs associated with galaxies have velocity extents wider than 80~\kms,
and are easily distinguishable from other FVWs.
FVW172.5+6.0 also show its excess emission over wide velocity extent
comparable to those of galaxies.
There is indeed a galaxy
named `LEDA 2126032' toward this direction according to the SIMBAD
astronomical database. The galaxy has a half-light radius of
$\sim$3 arcsec (2MASS Extended Source Catalog),
and it's radial velocity is not known.
If it is a dwarf galaxy having typical physical half-light radius of 100 pc, it
would be at 6~Mpc. Its association with FVW172.5+6.0 is possible.
No other FVWs appear to be external galaxies.
\clearpage
\begin{deluxetable}{lccrcc}
\tabletypesize{\scriptsize}
\tablewidth{0pt}
\tablecaption{FVWs Coincident with HVCs\label{hvcs}}
\tablehead{
\colhead{}    &  \multicolumn{5}{c}{HVC Parameters}\\
\cline{2-6} \\
\colhead{}           &
\colhead{($\ell, b$)}     & 
\colhead{Size}      & \colhead{Velocity}      &
\colhead{N$_{\rm H}$}  & \colhead{}\\
\colhead{FVW Name}           &
\colhead{(\degr)}      & 
\colhead{($\degr$)} & \colhead{(\kms)}      &
\colhead{($10^{18}$ cm$^{-2}$)} & \colhead{Reference}}
\startdata
FVW196.5+1.5 &(197.0, +1.5)&$\sim 1.0$ &$-$66.0&100&1\\
FVW203.0+6.5&(202.8, +6.4)&2.5&108.4&120&2\\
FVW215.0+7.0&(215.0, +7.0)&1.1&103.0&50&2\\

\enddata
\tablerefs{
(1) \citet{wakker01}; (2) \citet{wakker91}
.}
\end{deluxetable}
\clearpage
For comparison with HVCs, we used HVC catalogs of
\citet{wakker91} including about 560 HVCs of all sky,
and \citet{wakker01} containing summary catalog
based on the absorption line study for the HVCs and IVCs.
In addition, we compared the catalog of FVWs
with those of compact isolated HVC (CHVC) \citep{deheij02, putman02}.
\citet{deheij02} identified about 900 CHVCs
using LDS in the northern sky, and
\citet{putman02} identified
about 2000 compact and extended HVCs in southern sky
using the \schi Parkes All-Sky Survey (HIPASS),
which covers LSR velocity range between $-700$~\kms and $+1000$~\kms
with $15 \arcmin .5$ spatial resolution and 26~\kms velocity resolution.
As shown in Table~\ref{hvcs}, three FVWs are identified as HVCs.
Although $\sim$ 30\% of FVWs unrelated to the known sources
show isolated bumps in their spectra
(see Section 4.2),
most of them are not matched with HVCs in available HVC catalogs.

\section{Discussion}

We have identified 87 FVWs. 
Among them, 6, 4, and 3 FVWs are found to be coincident 
with SNRs, nearby galaxies, and HVCs, respectively. 
The rest (85\%) are not associated with 
any obvious objects that would be 
responsible for their large velocities. 
We already pointed out that FVWs associated with 
galaxies have large, easily-distinguishable velocity extents and
none of the rest FVWs, perhaps except FVW172.+6.0,
seems to be associated with external galaxies. 
Therefore, FVWs are thought to be Galactic objects. 
In this section, we discuss possible origins based on their 
observed properties.  

\subsection{Are they SNRs?}

KK04 argued that 
the number of observed SNRs with \schi shell (\schi SNRs)
is much less than that of the expected
and suggested that FVWs could be such missing old SNRs. 
In fact, in large-scale $(\ell,v)$ diagrams, 
the morphologies of most FVWs cannot be discriminated from 
the FVWs associated with known SNRs. 
And high-resolution observation showed that at least 
one FVW, FVW 190.2+1.1, is a rapidly ($\sim 80$~\kms) expanding \schi shell
that is most likely an old SNR remnant \citep{kks06}. 
Hence, old SNRs are considered to be primary candidates for FVWs.

The distribution of FVWs, however, appears 
significantly different from 
either that of the known SNRs or what we expect for old \schi SNRs. 
The known SNRs are strongly concentrated toward the inner Galaxy and 
toward the Galactic plane. If we use 265 SNRs in Green's 
catalog \citep{green06},
the longitude distribution 
has a maximum at $\ell\approx 10~\degr$ and a FWHM of $\sim 90~\degr$ while  
the latitude distribution is almost symmetric 
with a FWHM of $\sim 1\degr$. 
The expected distribution of old SNRs identifiable 
in \schi emission had been calculated by KK04 using a simple 
model where it was assumed that 
the Galaxy is composed of a uniform, axi-symmetric gaseous disk  
with a central hole and that only Type Ia supernovae produce isolated
SNRs which are exponentially distributed with a radial scale length of 3 kpc.
According to their result,
visible \schi SNRs are concentrated in the inner Galaxy 
along the loci of tangent points ($\ell\approx 40\degr$ and $320\degr$)
because the SN rate is higher in the inner Galaxy and 
because the systemic velocities of SNRs in those regions are close to
either the maximum or the minimum velocities of
the Galactic background emission. The number of 
visible \schi SNRs is $\sim 100$, 75\% of which is within $|\ell|\le 70\degr$.
KK04 did not calculate the latitude distribution, but 
if SN Ia distribution follows 
thin old stellar disk with an exponential scale height of $\sim 300$~pc
as in their model, 
the FWHM of these visible \schi SNRs would be $\sim 6$\degr. 
These results, showing that the expected distribution of visible 
\schi SNRs is quite different from 
that of FVWs, however, does not 
necessarily rule out the SN origin for FVWs:
Firstly, the number of \schi SNRs in the inner Galaxy 
could be significantly less than that in the model if 
the interstellar space there is largely filled with a very tenuous 
gas instead of warm neutral medium (0.14 cm$^{-3}$) 
as KK04 pointed out.
And, secondly, in the outer Galaxy, there could be more visible 
\schi SNRs than the model 
because the identification is relatively easy there.
Also core-collapse SNe may not be spatially and/or timely 
correlated in there, so that they result in \schi SNRs, too. 
However, the significantly larger number in the outer Galaxy and their large
scale height seem to indicate that not all of FVWs could be 
old SNRs. 

\begin{figure}
\plotone{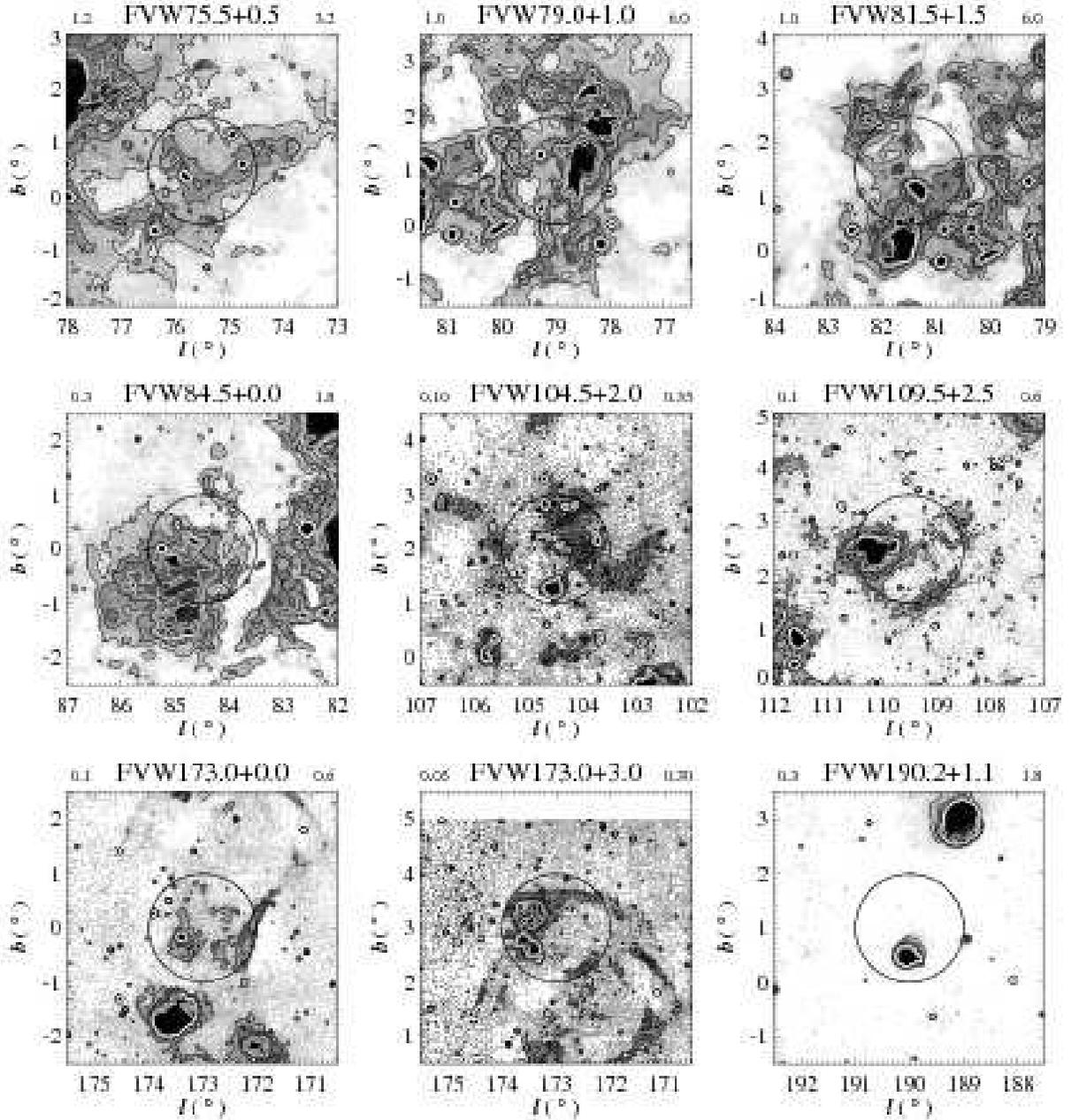}
\figcaption{
Effelsberg 11-cm continuum images of 9 FVWs
having noticable extended emission.
The minimum and the maximum contour levels in brightness temperature (K)
are written
at the top left and the top right corner of the image.
Contours are equally spaced.
The grey image and the contours are focused on showing faint emission.
The circle at the center represents area within diameter of $2\degr$.
\label{bonnfig}}
\end{figure}

If FVWs are SNRs, it may be possible to see faint radio continuum 
emission associated with them.
We have searched associated radio emission toward 38 FVWs
covered in the Effelsburg 11-cm continuum survey \citep{reich11, furst11}
($0\degr \la  \ell \la 240\degr$, $\mid b \mid \la 5\degr$).
9 FVWs are found to have noticable extended emission
within $2\degr$-sized circle in addition to
the 6 FVWs coincident with the known SNRs (Fig.~\ref{bonnfig}).
They are summarized in Table~\ref{contitable}.
The cases of FVW173.0+0.0 and FVW173.0+3.0 are interesting:
There is an about $3\degr$-size, incomplete shell-like continuum feature
centered at ($\ell, b$) $\sim(172~\degr .5 +1~\degr .5)$,
and \schi emission of FVW173.0+3.0 appears to be enclosed
by the filamentary continuum emission
while FVW173.0+0.0 lies on the south of the shell-like continuum.
Other than these two,
there is no obvious coincident shell-like continuum features.
In order to prove the association of these radio features with FVWs,
further studies are required.
\clearpage
\begin{deluxetable}{ll}
\tabletypesize{\scriptsize}
\tablewidth{0pt}
\tablecaption{FVWs with Coincident Radio Continuum Emission\label{contitable}}
\tablehead{
\colhead{FVW Name}&\colhead{Radio Continuum Characteristics}}
\startdata
FVW75.5+0.5 & Extended feature, not well matched with \schi\\
FVW79.0+1.0 & Complicated area\\
FVW81.5+1.5 & Complicated area\\
FVW84.5+0.0 & Complicated area\\
FVW104.5+2.0 & \schii region, not well matched with \schi \\
FVW109.5+2.5 & \schii region, not well matched with \schi \\
FVW173.0+0.0 & 3\degr-sized filamentary shell-like feature\\
FVW173.0+3.0 & Part of 3\degr-sized filamentary shell-like feature \\
FVW190.2+1.1 & \schii region, not well matched with \schi \\
\enddata
\end{deluxetable}
\clearpage

\subsection{Anomalous Velocity Clouds}

About 30\% of FVWs unrelated to known SNRs, galaxies, or HVCs show
isolated bumps in their spectra, e.g., FVW15.5$-$10.5 in Fig.~\ref{fvws}.
They appear as \schi 21 cm wings in
large-scale \lv diagrams because they are very close to the Galactic
\schi emission.
These FVWs therefore could be clouds with anomalous velocities
rather than expanding shells.
Table~\ref{isopeak} lists such FVWs.
The FVWs associated with known SNRs, HVCs, or galaxies are not included,
even though they show isolated bumps in their spectra. The bumps in some
FVWs in Table~\ref{isopeak}, e.g., FVW75.5+0.5, FVW293.0+7.9, and FVW312.3+3.6, are
weak ($\sim 1$~K) and not apparent in Fig.~\ref{fvws} (see the notes in
Table ~\ref{isopeak}). On the other hand, there are  FVWs that appear to
have very weak ($\sim 0.2$~K), broad bumps in Fig.~\ref{fvws}, e.g.,
FVW95.5+7.0 and FVW126.0$-$1.0, but they are hardly identified as
isolated bumps in their spectra and not included in Table~\ref{isopeak}.

We have determined central LSR velocity, FWHM, and brightness temperature
of FVWs with isolated bumps 
by fitting their spectra
into a Gaussian plus a second order polynomial function.
Table~\ref{isopeak} summarizes the result.
Their central velocities vary from \vlsr$=-124.3$ to $+170.3$~\kms. 
Their FWHMs are 7.0--25.5~\kms with median value of 12.4~\kms.
Their median peak brightness temperature is 1.0~K,
and the median hydrogen column density is $2.4 \times 10^{19}$~cm$^{-2}$.
Table~\ref{isopeak} also lists the deviation velocity,
which is defined as the excess of central velocity over
the minimum or maximum velocities
permitted by the Galactic rotation \citep{wakker91a}.
Here, we assume a Galactic model that \citet{wakker91a} used, i.e.,
a 4~kpc thick disk with radius 26~kpc and a 
flat rotation curve with a value of $v_\odot=220$~\kms and $R_\odot=8.5$~kpc.
The deviation velocities range from $-33.8$ to +46.6~\kms,
and the median value is 31.9~\kms.
The negative \vdev means that the centeral velocity is
within the allowed velocity range,
which implies that the used disk model does not describe well 
the real velociy boundaries.

The velocities of FVWs indicate that 
they could have been classified either as HVCs or 
IVCs. 
HVCs are generally defined as clouds with velocities greater 
than 90~\kms\ with 
respect to the local standard of rest (LSR) \citep{wakker01, putman02}. 
At low galactic latitudes, however, 
this definition is not very useful and 
an alternative definition in terms of the deviation velocity could be 
used, i.e., \vdev$\ga$ 50--60~\kms \citep{wakker91a, putman02}.
Observationally IVCs are just lower velocity counterparts of HVCs,
which are defined as clouds with 30--40$\la|v_{\rm LSR}|\la 90$~\kms\ 
\citep{wakker01, richter03}
in terms of LSR velocity.
About one third of the FVWs in Table~\ref{isopeak} have central LSR velocities 
$|v_{\rm LSR}|=100$ to 170~\kms, so that they may be 
classified as HVCs while the rest as IVCs. 
On the other hand, all of them have deviation velocities less than 
50~\kms, which puts them in the category of IVCs.  
An interesting feature of HVCs is that they show a systematic
velocity change, i.e., most HVCs are at positive velocities
between $\ell \sim 200\degr$ and $320\degr$, while most of them are at
negative velocities in the rest \citep{wakker91}. This asymmetric
velocity structure was explained by a model where HVCs are debris of
the parental cloud of the Local Group,
falling toward the center of the Local Group
\citep{blitz99, deheij02m}. The FVWs in Table ~\ref{isopeak} show somewhat
opposite velocity trend, i.e., about 70\% of FVWs between
$\ell \sim 200\degr$ and $320\degr$ are at negative velocities,
which is similar to the trend of the entire FVWs (\S~3.2). This makes it
difficult to consider the FVWs as a low-velocity population of HVCs.

It is also possible that the FVWs with isolated bumps
are small, fast-moving clouds unresolved in the survey.
Recently, small ($\la 10$~pc), fast-moving clouds are revealed
by high-resolution observations 
in the disk/halo interface region of the inner Galaxy, so called halo clouds
\citep{lockman02l}, in the disk ($|b|<1.^\circ3$) of the inner Galaxy \citep{stil}, 
and in the anti-center region \citep{stani06}.
It was suggested that our Galaxy may possess a population of those clouds
distributed both throughout the disk and up into the halo
rotating with the Galactic disk,
and that about one-half of the Galactic \schi halo
could be in the form of discrete clouds \citep{lockman02l,stil}.
They could be clouds formed in Galactic fountain flows
or infalling intergalactic material in the on-going construction of the Galaxy
\citep[][and references therein]{stani06}.
The observational properties of 
the halo clouds and the clouds in the inner Galactic disk and in the anti-center
region are summarized by \cite{stani06} (their Table 3).
The properties of FVWs in Table.~\ref{isopeak} are quite comparable to 
those of halo clouds, i.e., their velocity dispersion (12~\kms) and 
\schi column densities (2$\times 10^{19}$~cm$^{-2}$) are similar.
The fast-moving clouds in the inner Galactic disk 
have a factor of 2 smaller FWHM and much larger \schi column densities, while 
the clouds in the anti-center region have comparable 
\schi column densities but a factor of 3 smaller FWHM.
But, since FVWs are not resolved, this comparison should be 
considered as provisional until high-resolution maps of FVWs are 
available.  

We should keep in mind that the isolated bump in spectrum
does not necessarily mean that a FVW is not part of a SNR.
Because the ISM is not uniform, the accelarated \schi shells of SNRs
often show bump-like features in their \schi spectra \citep[e. g.,][]{koo90}.
It is also worth to note that the fast-moving small clouds 
usually appear related to diffuse filamentary structures 
\citep{stani06}, which suggests a possibility of disrupted SNR shells.
The nature of those cloud-like FVWs needs further studies.
\clearpage
\begin{deluxetable}{lrrrrrl}
\tablecolumns{7}
\tabletypesize{\scriptsize}
\tablewidth{0pt}
\tablecaption{
FVWs with Isolated Bumps in Position-Velocity Maps
\label{isopeak}}
\tablehead{
\colhead{}           &
\colhead{$v_{\rm center}$\tablenotemark{a} }           &
\colhead{$v_{\rm dev}$\tablenotemark{b} }           &
\colhead{FWHM\tablenotemark{a}}      &
\colhead{Peak T$_{b}$\tablenotemark{a}}	&
\colhead{N$_{\rm HI}$\tablenotemark{c}} & \colhead{}\\
\colhead{FVW Name}	&          
\colhead{( \kms )}           & \colhead{( \kms )}      &
\colhead{( \kms )}           & \colhead{( K )}	&
\colhead{( 10$^{18}$ cm$^{-2}$ )} & \colhead{Note}}
\startdata

FVW10.0+7.5& $-$43.6&  20.3&  16.6&   1.5&  50.3&1\\
FVW15.5$-$10.5& $-$34.1&  12.4&  13.8&   0.7&  19.9&1\\
FVW18.0$-$6.5& $-3$9.3&  $-$6.2&  12.9&   0.9&  23.4&1\\
FVW71.0$-$4.0&$-$105.9& $-$33.8&   7.2&   2.5&  35.5&1\\
FVW75.5+0.5&  38.4&  31.4&   8.1&   0.9&  13.6&2\\
FVW84.5+0.0&  45.6&  44.5&  12.2&   2.8&  65.8&1\\
FVW86.0$-$9.5&  46.9&  46.4&   5.0&   1.8&  17.8&3\\
FVW201.0+9.5&  72.3&  20.0&  17.0&   0.4&  14.4&1\\
FVW252.5+10.5& $-$34.4&  34.4&   8.9&   0.4&   6.3&1\\
FVW260.9$-$1.9& 170.3&  24.2&  25.5&   1.6&  82.5&1\\
FVW261.6+2.3& 164.0&  17.6&  19.1&   0.9&  32.7&1\\
FVW283.4+3.0& $-$47.3&  41.3&   7.0&   0.9&  11.7&2\\
FVW291.0+2.9& $-$61.2&  46.6&  11.2&   2.4&  52.2&1\\
FVW293.0+7.9& $-$56.7&  39.4&  10.1&   1.0&  20.7&2\\
FVW298.3$-$3.7& 154.5&  24.4&  11.2&   0.6&  13.3&4\\
FVW304.5$-$5.7& $-$76.6&  38.1&  14.1&   1.9&  51.6&1\\
FVW311.9$-$1.1&$-$100.8&  44.6&  12.3&   0.6&  14.5&1\\
FVW312.3+3.6& $-$82.3&  25.2&  12.5&   1.2&  30.4&2\\
FVW313.4+0.6&$-$102.6&  42.5&  12.4&   1.0&  24.2&1\\
FVW324.5+4.2&$-1$24.3&  32.3&  22.4&   0.9&  39.3&1\\
\tableline
Median Values& $-41.5$&  31.9 &  12.4 &  1.0&  23.8\\
\enddata
\tablenotetext{a}{$v_{\rm center}$, FWHM, and peak T$_{b}$ were derived by
Gaussian fitting.}
\tablenotetext{b}{$v_{\rm dev}$ is defined as the excess of
central velocities over the minimum and maximum velocities
permitted by the Galactic differential rotation.
Thus, nagative \vdev represents $v_{\rm center}$ is within the velocity range
permitted by the given disk model.}
\tablenotetext{c}{N$_{\rm HI}$ is derived from
the total integrated intensity of Gaussian component.}
\tablecomments{ (1) It is a well defined, isolated clump.
(2) The excess emission is weak in comparison to the Galactic HI, so
that the spectral bump is not apparent in Fig.~\ref{fvws},
although it is clear in the spectrum.
(3) As seen in Fig.~\ref{fvws},
FVW$86.0-9.5$ consists of smoothly extended \schi emission and small clump
at higher velocity. The parameters listed in this table are
the Gaussian fitting results of the small clump.
(4) It shows double bumps in spectrum. The parameters are
those of the larger bump.}
\end{deluxetable}
\clearpage
\subsection{Other Possibilities}

Another candidate for FVWs is stellar winds and wind-blown shells.
Low- and high-mass protostars produce fast expanding winds or jets.
OB stars or Wolf-Rayet (WR) stars eject strong winds forming expanding ionized or neutral \schi shells.
AGB stars also expel their energies into the surrounding medium
as slow and fast winds.
Here, we will discuss if they could be detectable in the \schi surveys
used in this work.

Low- and high-mass protostars
lose their masses by stellar winds and jets.
The wind could be neutral intrinsically \citep{lizano88}
or neutral hydrogen could be produced by
dissociation of molecular hydrogen in shocks \citep{bally83}.
Neutral atomic stellar wind was detected toward
young stellar object HH 7-11 at 350 pc from the sun \citep{lizano88}.
The average antenna temperature of neutral wind of the HH 7-11
is $\sim 0.01$~K when it is observed at the Arecibo telescope.
This corresponds to an antenna temperature of $\la 0.001$~K
for a 25-m or 64-m telescope.
Therefore, if the observed parameters of HH 7-11 represent typical wind parameters
of low-mass protostars, FVWs identified in this paper ($T_A\ga 0.1$~K) 
cannot be due to such neutral winds.

Neutral atomic stellar wind has been detected toward high-mass protostars, too.
\citet{russell92} detected 
atomic stellar wind expanding at $\sim 90$~\kms toward DR 21,
which contains one of the most powerful outflow sources in the Galaxy.
The \schi mass of the wind is 24~\ms at the
generally accepted distance of DR 21 (3~kpc),
which is three orders of magnitude more massive than
the \schi wind seen in HH 7-11.
This would yield $\sim 0.007$K for a 25 m telescope.
If we assume the observed parameters of DR 21 is typical,
the neutral winds from massive protostars would not be detectable
unless they are closer than $\la 1$~kpc.

Strong stellar wind from early type OB stars or WR stars is likely to be
an another possible source producing fast-moving \schi gas.
The mechanical luminosity of winds from OB stars range from 
10$^{31}$~erg s$^{-1}$ to $10^{37}$~erg s$^{-1}$ with a 
typical expansion velocity of 1,500~\kms \citep{bieging90}.
This wind sweeps up ambient medium into a shell
which could be observed as FVWs.
Most observed \schi shells
around OB stars, however, have low expansion velocities, i.e., $\la 20$~\kms
\citep[e.g.,][]{cappaher00, cappapi02, boulva99}.
But these expansion velocities might be lower limits
because their endcaps are not usually detected.
According to \citet{weaver77},
the radius of expanding shell is given by
$R(t)=26 n_0^{-0.2} L_{36}^{0.2} t_6^{0.6}$~pc,
where $n_0$ is ambient atomic density of H nuclei in \cmc, 
$L_{36}$ is the wind luminosity in $10^{36}$~erg s$^{-1}$,
and $t_6$ is the age in $10^6$~yr.
From this, we may express the mass in the shell $M(t)$ in 
terms of its expansion velocity $v(t)$; 
$M(t)=\frac{4 \pi}{3} \rho_0 R(t)^3 $
$=0.4 n_0^{-0.5} L_{36}^{1.5} {v_2}^{-4.5}$~\ms,
where $v_2$ is expansion velocity in 100~\kms.
When this shell is observed with a 25-m telescope,
it would yield a
brightness temperature of $T_b \simeq 0.6 M(t) d^{-2} \Delta v^{-1}$,
where M(t) is in ~\ms, d is distance in kpc, and $\Delta v$ is velocity
interval over which the \schi shell would be detected in ~\kms. Since 0.1~K is
a detection limit in this work, the ones at distance less than
$d \la 0.11 n_0^{-0.25} L_{36}^{0.75} v_2^{-2.75}$~kpc
could be detected assuming $\Delta v \simeq 2 v(t)$.
If we assume that the shells of expansion velocity $v(t) \ga 50$~\kms
are detectable, then the distance should be less than
$d \la 0.7 n_0^{-0.25} L_{36}^{0.75}$~kpc.
Thus, nearyby ($\la 4$~kpc) wind shells with large wind luminosity
($\ga 10^{37}$~erg s$^{-1}$) can be detected, assuming $n_0 \simeq 1$~\cmc.

We checked if O-type or Wolf-Rayet stars having large wind luminosities
are in the region of FVWs,
using the Galactic O star catalog of \citet{garmany82}
and the 7th catalog of Galactic Wolf-Rayet stars of \citet{vanderh01}.
About 15~\% of FVWs unrelated to the known sources
have more than one stars in 1~\degr-radius area.
They are summarized in Table~\ref{owrstars}.
12 and 3 FVWs have nearby O stars and WR stars,
respectively.
All 3 FVWs coincident with WR stars are coincident with O stars, too.
They are at quite close distance ($d \la 3$~kpc).
Thus, some of those FVWs in Table~\ref{owrstars}
could be caused by winds from massive stars.
However,
it is worthy to note that the 
FWHM of the latitude distribution of 
early type stars ($\sim 1$\degr) \citep{pal03}
is much smaller than that of FVWs ($\sim 13$~\degr). 
\clearpage
\begin{deluxetable}{lccllcc}
\tablewidth{0pt}
\tabletypesize{\scriptsize}
\tablecaption{FVWs Coincident with Galactic O-type or WR stars\label{owrstars}}
\tablehead{
\colhead{}    &\colhead{}&  \multicolumn{4}{c}{Parameters of The Nearest Star}\\
\cline{3-7} \\
\colhead{}           & \colhead{Number}           &
\colhead{($\ell, b$)}     & 
\colhead{}   &\colhead{}   & \colhead{Angular Distance}      &
\colhead{Distance}  \\
\colhead{FVW Name}           &\colhead{of Stars\tablenotemark{a}} &
\colhead{(\degr)}      & \colhead{Other Names} &\colhead{Spectral Type}    &
\colhead{(\degr)} & \colhead{(kpc)}}
\startdata
\multicolumn{7}{c}{Galactic O stars}\\
\tableline
FVW6.5-2.5&           2&(  6.9, $-$2.1)&HD        165921 &O7.5V  &0.58& 1.3\\
FVW15.5-10.5&           1&( 15.3,$-$10.6)&HD        175876&O6.5III  &0.22& 2.3\\
FVW75.5+0.5&           6&( 75.9, +0.8)&HD        193682 & O5. V &0.50& 1.8\\
FVW79.0+1.0&           4&( 79.0, +1.3)&BD      +40 4179 &O8. V  &0.30& 1.2\\
FVW83.5+4.0&           1&( 83.8, +3.3)&BD      +45 3216 &O8. &0.77& 1.9\\
FVW104.5+2.0&           1&(103.8, +2.6)&HD        210839 & O6. I &0.92& 0.8\\
FVW109.5+2.5&           5&(109.6, +2.7)&HD        216532 & O9.5V &0.21& 0.9\\
FVW173.0+0.0&           4&(173.0, +0.1)&HD         35619 & O7. V &0.10& 3.2\\
FVW173.0+3.0&           3&(173.5, +3.2)&HD         37737 & O9.5III &0.54& 1.3\\
FVW190.2+1.1\tablenotemark{b}&           1&(190.0, +0.5)&HD         42088 &O6.5V &0.50& 1.5\\
FVW313.4+0.6&           3&(313.5, +0.2)&HD        125241 & O8.5I &0.47& 3.2\\
FVW342.5+1.7&           9&(342.8, +1.7)&HD        151515 & O7. III &0.30& 1.9\\
\tableline
\multicolumn{7}{c}{Galactic WR stars} \\
\tableline
FVW75.5+0.5&           3&( 75.7, + 0.3)& Sand 5&WO2  &0.30& 0.9\\
FVW79.0+1.0&           2&( 79.7, +0.7)& V1923 Cyg, AS 422 &WN7/WCE+?& 0.77 &\\
FVW342.5+1.7&           1&(343.2, +1.4)& HD 151932 &WN7h&0.77& 2.0\\

\enddata
\tablecomments{All parameters 
of the nearest star in Table ~\ref{owrstars}
are from \citet{garmany82} or \citet{vanderh01}
except angular distance,
which simply indicates angular distance of the nearest star from a FVW center.
Distance of HD 151932 leaves blank since it is not noted in the reference catalog.}

\tablenotetext{a}{Number of stars in 1~\degr  radius. }
\tablenotetext{b}{High resolution study of \citet{kks06} revealed that
FVW190.2+1.1 is possibly old SNR and not quite related with HD 42088.}
\end{deluxetable}
\clearpage

Highly evolved cool stars in asymptotic giant branch (AGB) eject
strong stellar winds too and could be an another candidate 
for FVWs. 
For example, molecular outflow at a speed of $\sim 200$~\kms
has been detected in AGB carbon star V Hya \citep{knapp97}.
These molecular winds may collide with the interstellar medium
to produce fast-moving atomic gas.
In particular, the FWHM of the latitude distribution of 
carbon stars is comparable to that of FVWs, 
e.g., $\sim 10\degr$
based on the catalog of Galactic Carbon stars of \citet{alk01}.
But the mass loss rate of AGB stars is
low ($\sim 10^{-7}$ \ms yr$^{-1}$), so that 
the resulting 21-cm emission 
might be too faint.

In summary,
while some of FVWs could be produced by nearby massive stars with large wind
luminosity,
the stellar winds or wind-blown shells from
low- or high-mass protostars, or AGB stars
don't seem to be capable of producing most of FVWs.

\section{Conclusion}

FVW appears as a faint extended wing-like feature in large-scale \lv diagram.
Since they have excessive emission
at high velocities forbidden by the Galactic rotation,
there should be some dynamical phenomena associated. 
Thus, FVWs are closely related to the dynamical energy sources in the Galaxy
and their effects on the ISM.

In this paper, we have identified 87 FVWs
in the Galactic plane in use of the LDS and the SGPS data.
FVWs show double-peak distribution in longitude.
They have wide FWHM of about $13\degr$ in latitude
and slightly skewed to positive latitude.
About 70~\% of FVWs do have positive LSR velocities. 
These statistical properties appear to reflect the 
identifiability of FVWs in the first place, e.g., how smooth and sharp 
the boundary of the general Galactic \schi emission is in 
$(\ell,v)$ diagram. But there are properties that appear to be intrinsic too, e.g., 
the small number of FVWs at $\ell=130^\circ$--180$^\circ$. 
The column densities of most FVWs are less than $8 \times 10^{19}$ cm$^{-2}$.
We compared the catalog of FVWs with those of SNRs, galaxies, and HVCs,
and found that $\sim 85$~\% are not coincident with those known objects.

We investigated the possibility
that FVWs could be the old \schi SNRs invisible in radio continuum (KK04).
The distribution of FVWs is different from
that of the expected \schi SNRs,
and the FWHM of FVWs is wider than that of SNRs.
The possibility of FVWs to be the old \schi SNRs
is not promising in this regard.
Instead a good fraction of FVWs
may be related to anomalous velocity clouds or O-type/WR stars:
About 30\% of FVWs unrelated to the known sources have discrete \schi bumps 
that are a bit seperated from the Galactic \schi.
Those FVWs could be isolated clouds similar to
the newly discovered halo clouds \citep{lockman02l}.
About 15\% of FVWs unrelated to the known sources
have nearby O-type stars.
Thus, these FVWs could be due to the stellar winds from those stars.
The origin of the remaining 55\% of FVWs unrelated to the known sources
is uncertain.
In a sense that
the \schi SNRs are, at least, strong and fast enough to be detected as FVWs,
SNRs still remain as a strong candidate of FVWs.
If FVWs are indeed the oldest type of SNR, that which is essentially invisible
except via its \schi line emission, then only 
high resolution \schi observations might be able to 
reveal their nature.

\acknowledgments

\begin{figure}
\epsscale{.80}
\plotone{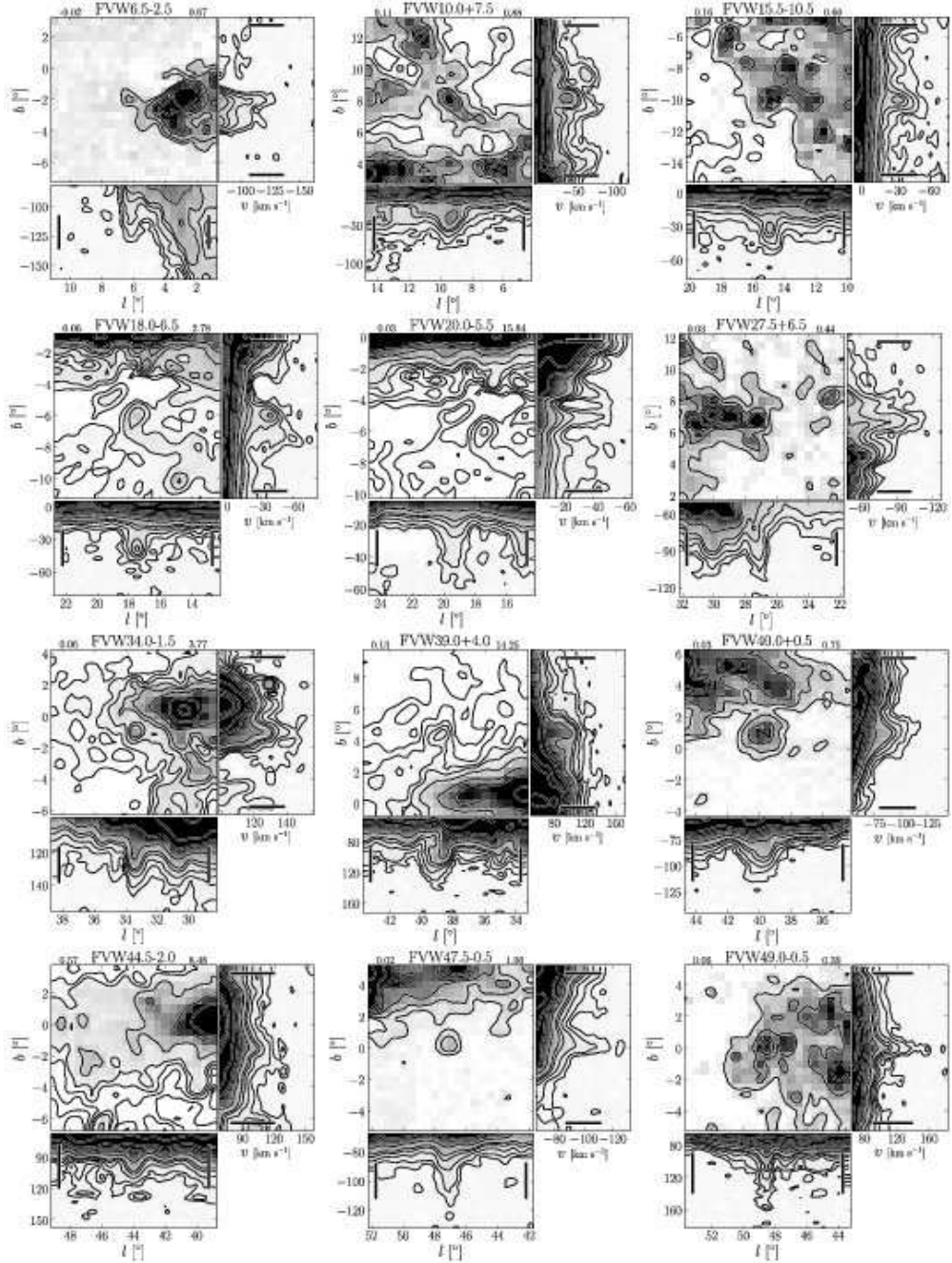}
\figcaption{
FVW's integrated image and its ($\ell$-$v$) and ($b$-$v$) diagrams
are presented at the center, bottom, and right.
The integrated velocity range is shown as
thick black lines in its position-velocity diagrams.
The minimum and maximum values of contours in the integrated map in K~\kms
are noted on the left and right side of the FVW name at the top.
($\ell$-$v$) and ($b$-$v$) diagrams are cut at the longitude or the latitude
of the FVW name
except FVW109.5+2.5 and FVW197.0+7.0
for which the slicing latitudes are marked in the figure.
The contour levels of position-velocity diagrams are
0.1, 0.2, 0.4, 0.7, 1, 2, 5, 10, 30, 50, 100~K in brightness temperature.\label{fvws}}
\end{figure}

We are very grateful to Jae-Jun Lee for his help with preparing Fig.~\ref{fvws}.
We would like to thank Naomi Melissa McClure-Griffiths
for providing the SGPS data.
We would like to thank Jay Lockman, Myung Gyoon Lee, and Ho-Seung Hwang
for their valuable discussions about other possibilities of FVWs.
We also thank to the anonymous referee for his/her comments that improved
the presentation of this paper.
This work was supported by the Korea Science
and Engineering Foundation (ABRL 3345-20031017). J.-h. K. has been supported
in part by the BK 21 program.

\clearpage

\clearpage
\epsscale{.80}
{\plotone{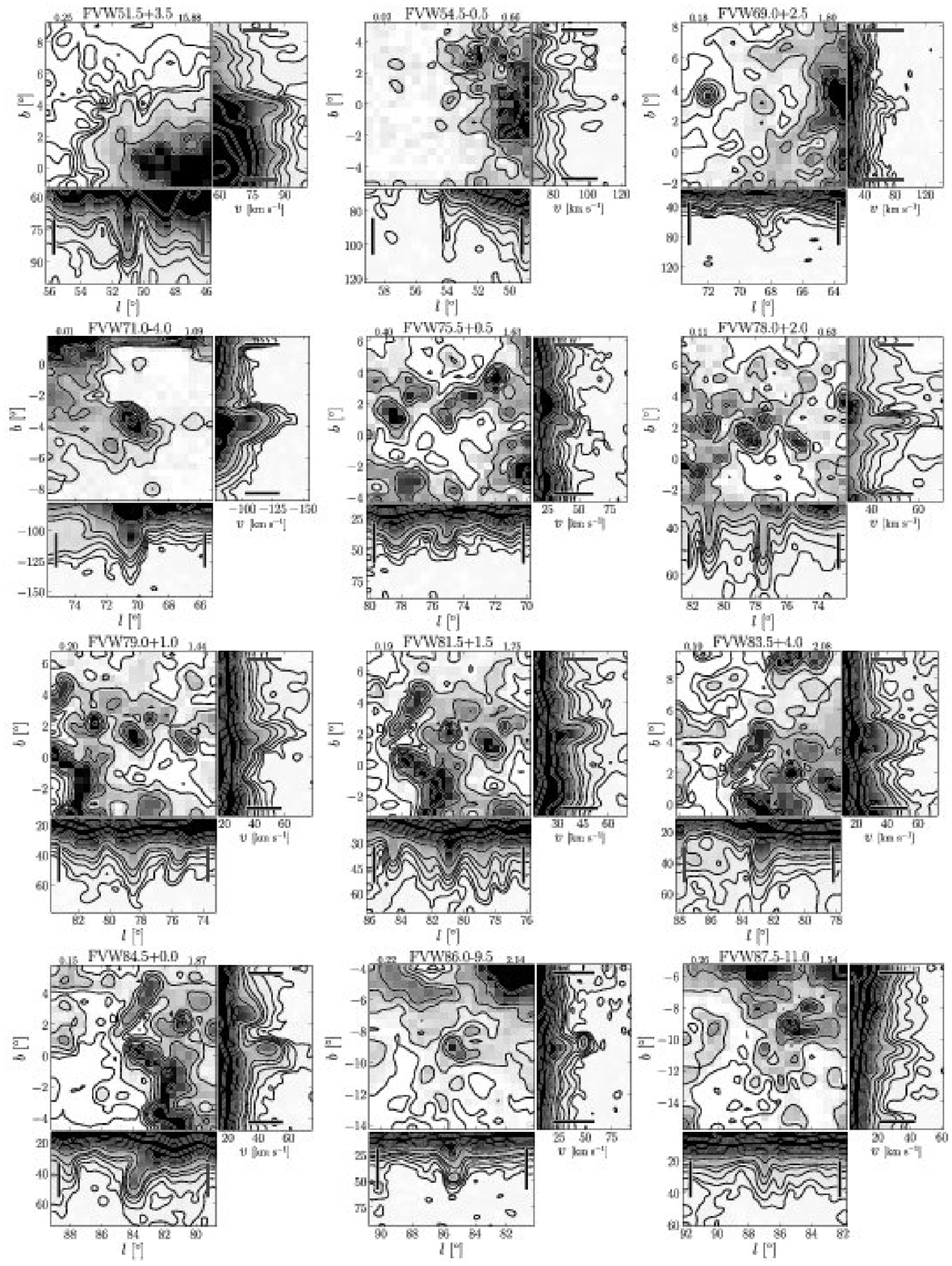}}
\clearpage
{\plotone{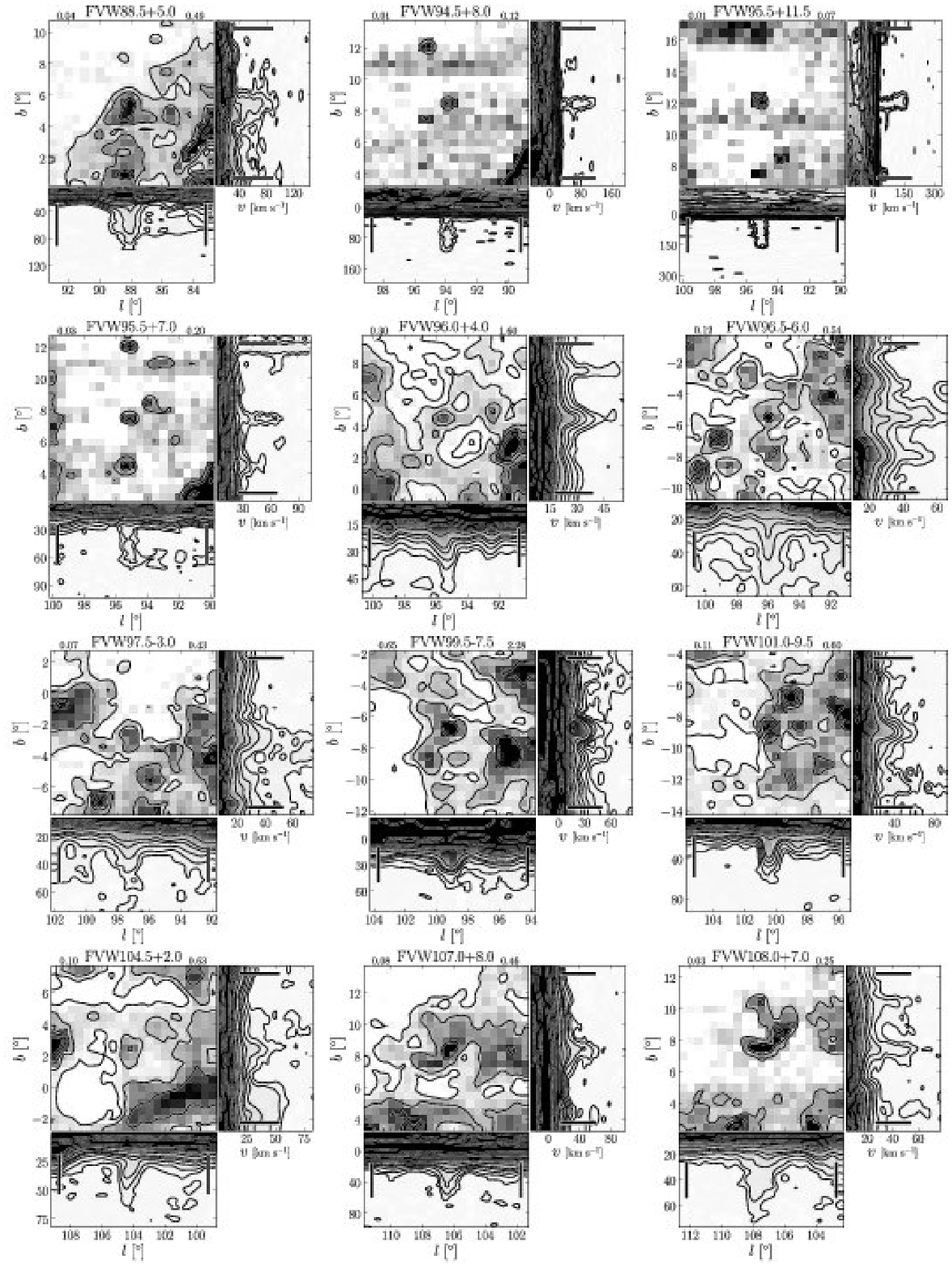}}
\clearpage
{\plotone{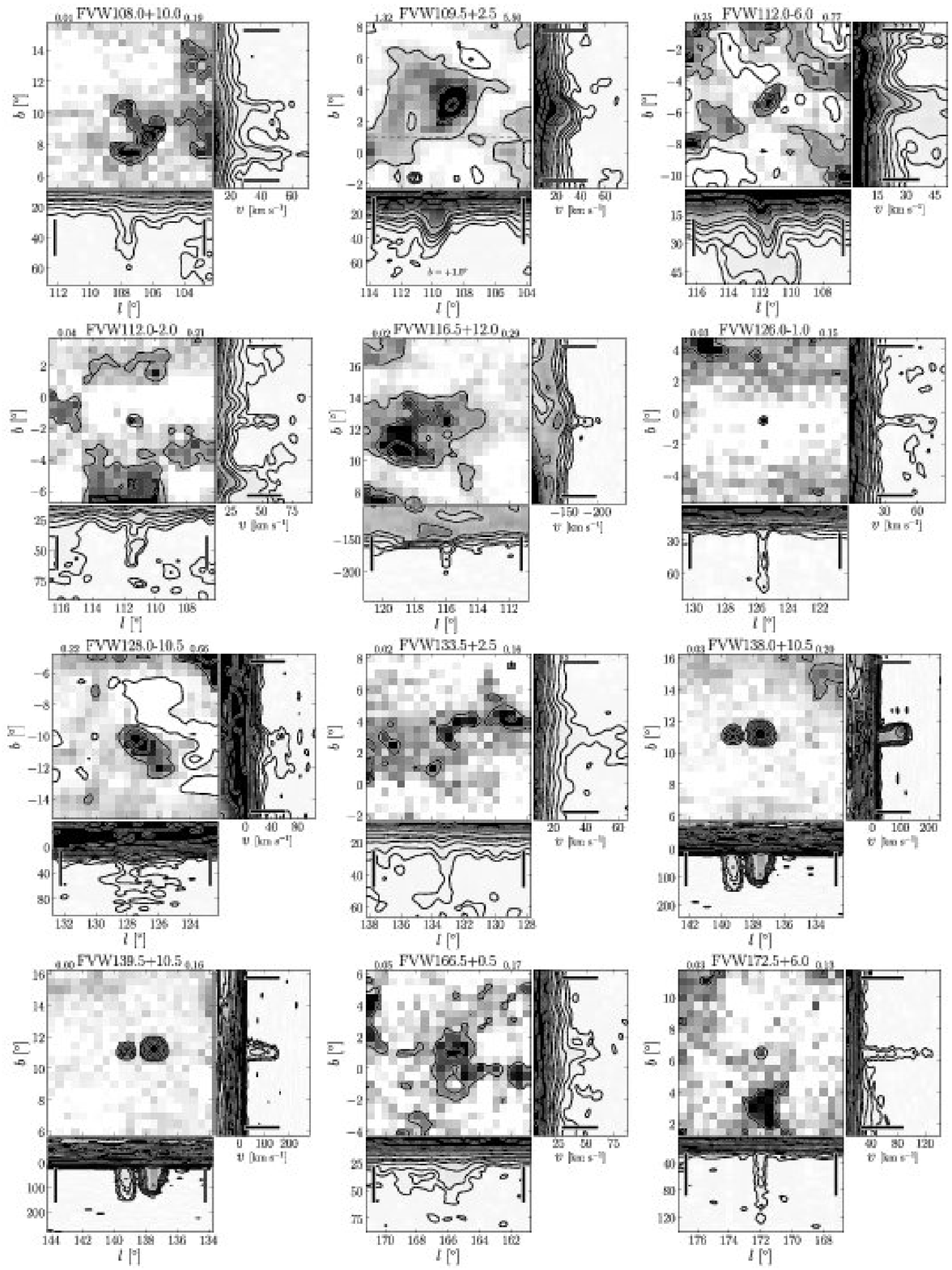}}
\clearpage
{\plotone{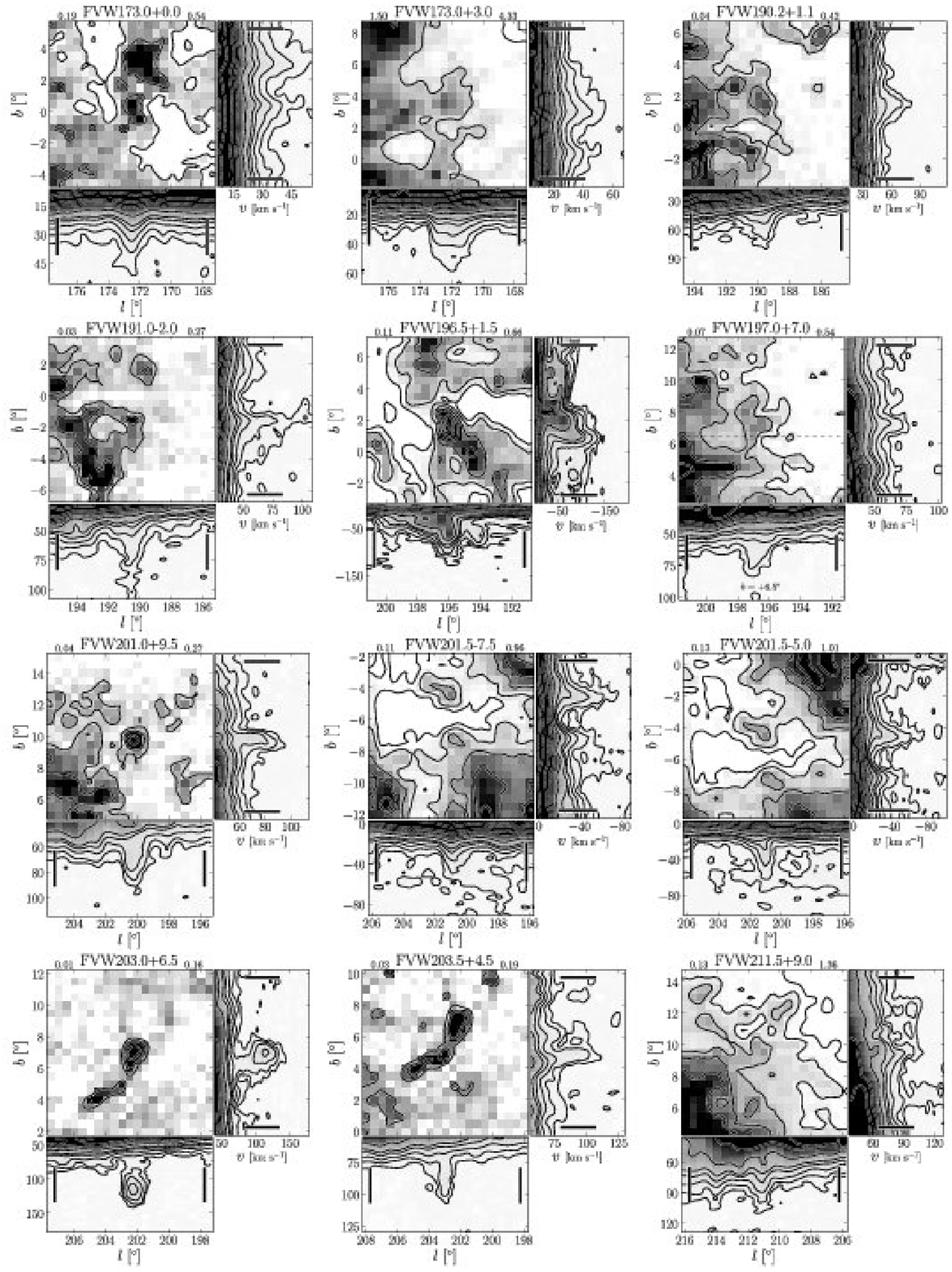}}
\clearpage
{\plotone{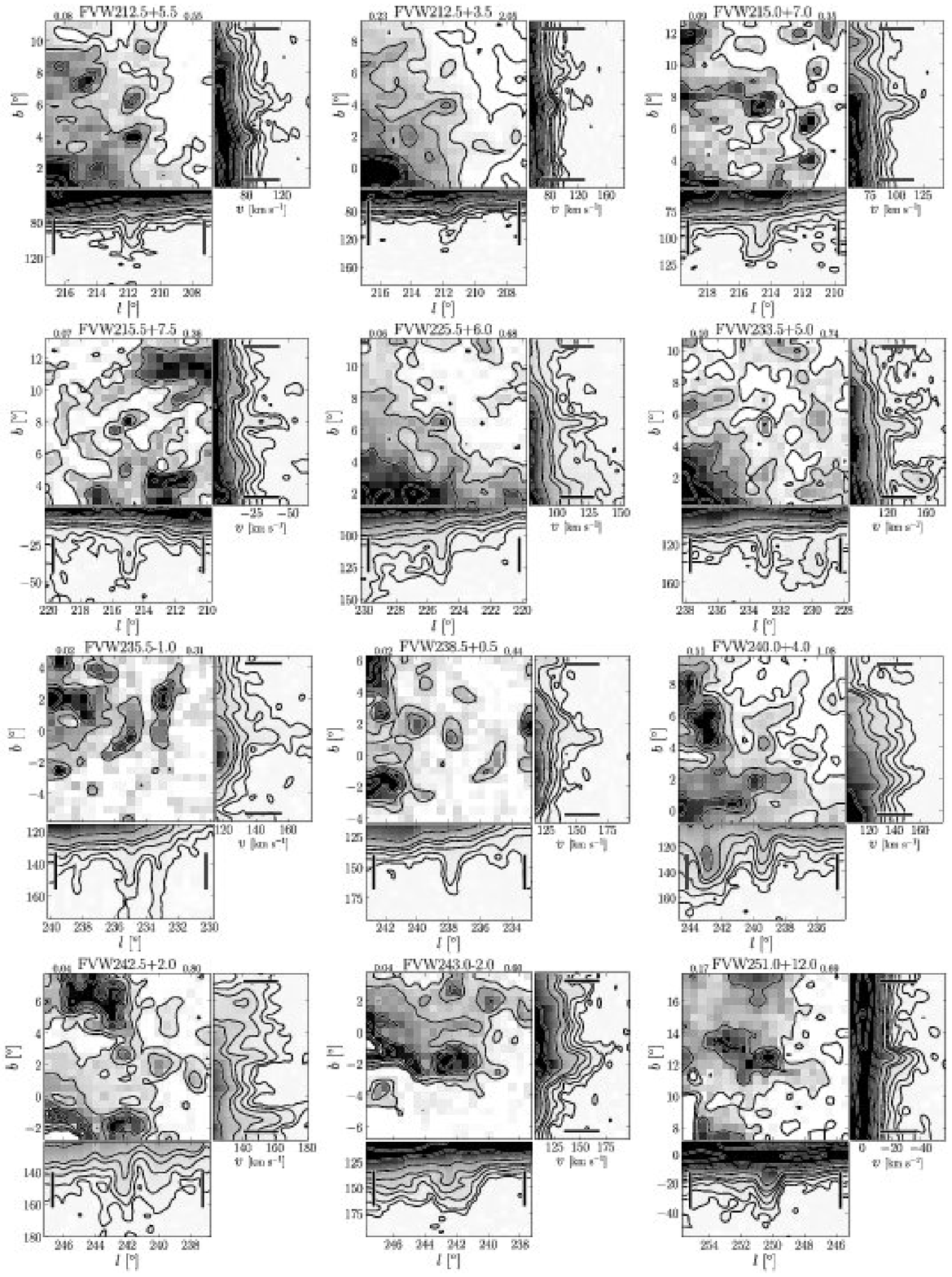}}
\clearpage
{\plotone{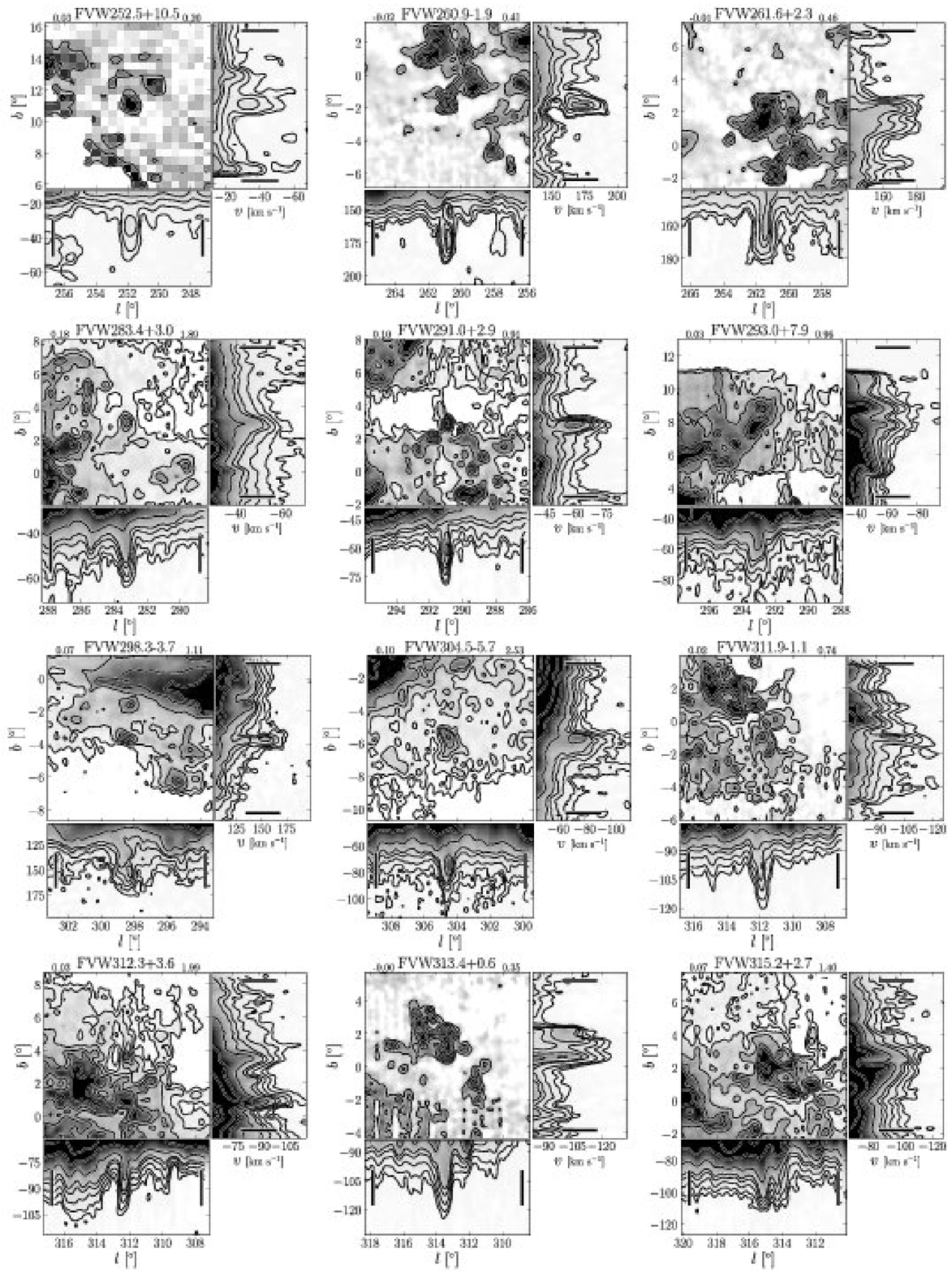}}
\clearpage
{\plotone{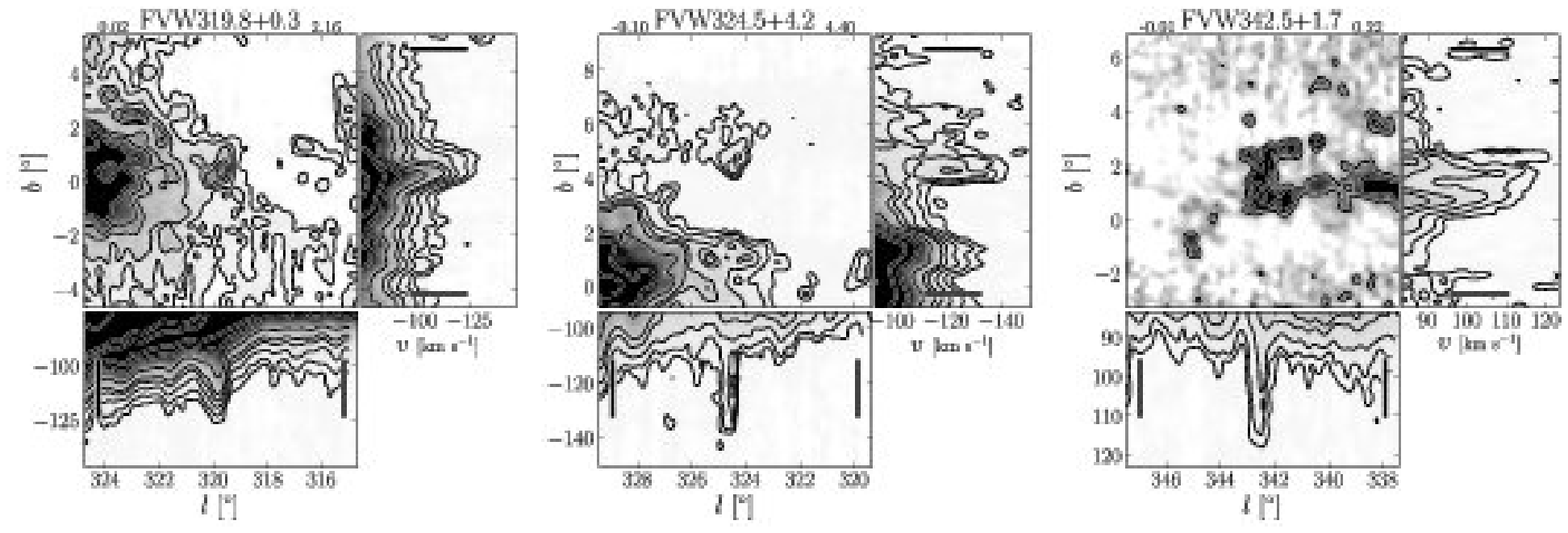}}
\clearpage

\end{document}